\documentclass[twocolumn,twocolappendix]{aastex63}

\submitjournal{the ApJ}

\usepackage{amsmath,amssymb,amsbsy,amsfonts}
\usepackage[british]{babel}
\usepackage{mdwmath}
\usepackage{color,xcolor}
\usepackage{url}
\usepackage{mathrsfs}
\usepackage{array}
\usepackage{subfigure} 
\usepackage{graphicx}
\usepackage{enumerate}

\newcommand{\eq}[1]{Eq.\,\eqref{#1}}
\newcommand{\eqs}[1]{Eqs.\,\eqref{#1}}
\newcommand{\cc}{\mathrm{c}}
\newcommand{\co}{\mathrm{const.}}
\newcommand{\ud}{\mathrm{d}}

\newcommand{\G}{\mathrm{G}}
\newcommand{\s}[1]{{\text{\tiny $#1$ }}\hspace{-1.5pt}}

\renewcommand{\theequation}{\thesection.\arabic{equation}}

\shorttitle{Choked accretion onto a Schwarzschild black hole}
\shortauthors{Tejeda, Aguayo-Ortiz \& Hernandez}

\begin{document}

\title{Choked accretion onto a Schwarzschild black hole: \\A hydrodynamical 
jet-launching mechanism}

\correspondingauthor{Emilio Tejeda}
\email{emilio.tejeda@conacyt.mx, aaguayo@astro.unam.mx, xavier@astro.unam.mx}

\author{Emilio Tejeda}
\affiliation{C\'atedras Conacyt --  Instituto de F\'isica y Matem\'aticas, 
Universidad Michoacana 
de San Nicol\'as de Hidalgo, Edificio C-3, Ciudad Universitaria, 58040 Morelia, 
Michoac\'an, Mexico}

\author{Alejandro Aguayo-Ortiz}
\affiliation{Instituto de Astronom\'ia, Universidad Nacional Aut\'onoma de 
M\'exico, AP 70-264, 04510 Ciudad de M\'exico, Mexico}

\author{X.~Hernandez} 
\affiliation{Instituto de Astronom\'ia, Universidad Nacional Aut\'onoma de 
M\'exico, AP 70-264, 04510 Ciudad de M\'exico, Mexico}

\begin{abstract}
We present a novel, relativistic accretion model for accretion onto a 
Schwarzschild black hole. This consists of a purely hydrodynamical mechanism in 
which, by breaking spherical symmetry, a radially accreting flow transitions 
into an inflow-outflow configuration. The spherical symmetry is broken by 
considering that the accreted material is more concentrated on an equatorial 
belt, leaving the polar regions relatively under-dense. What we have found is a 
flux-limited accretion regime in which, for a sufficiently large accretion 
rate, the incoming material {\it chokes} at a gravitational bottleneck and the 
excess flux is redirected by the density gradient as a bipolar outflow.  The 
threshold value at which the accreting material chokes is of the order of the 
mass accretion rate found in the spherically symmetric case studied by Bondi 
and Michel. We describe the choked accretion mechanism first in terms of a 
general relativistic, analytic toy model based on the assumption of an 
ultrarelativistic stiff fluid. We then relax this approximation and, by means 
of numerical simulations, show that this mechanism can operate also for general 
polytropic fluids. Interestingly, the qualitative inflow-outflow morphology 
obtained appears as a generic result of the proposed symmetry break, across 
analytic and numeric results covering both the Newtonian and relativistic 
regimes. The qualitative change in the resulting steady state flow 
configuration appears even for a very small equatorial to polar density 
contrast ($\sim$0.1\%) in the accretion profile. Finally, we discuss the 
applicability of this model as a jet-launching mechanism in different 
astrophysical settings.
\end{abstract}


\section{Introduction}
\label{s1}

Astrophysical jets are found in vastly different scenarios: from the parsec 
scales of the H-H objects associated with young stellar systems 
\citep{hartigan2009}, to the megaparsec scales of the radio lobes that 
accompany some radio galaxies and other Active Galactic Nuclei (AGN) 
\citep{beckmann12}. They are also inferred in connection with high energy 
phenomena such as long Gamma Ray Bursts (GRBs) following the collapse of a 
massive star \citep{woosley06}, jetted emission associated with micro-quasars in 
some X-ray binaries \citep{MR94}, X-ray flares after a stellar tidal disruption 
event \citep{burrows}, and short GRBs accompanying the kilonova explosion after 
the merger of two neutron stars \citep{GW170817}.

In recent decades, substantial progress has been made in understanding 
different aspects of astrophysical jets, particularly in relation to their  
acceleration and collimation \citep[see e.g.][]{qian2018,liska2019}. However, 
open questions remain concerning the process of launching the jet in the first 
place, as well as the details connecting the accreted and ejected flows 
\citep{romero2017}.

 \begin{figure*}
 \begin{center}
  \includegraphics[trim=0 55 0 55, clip, width=0.8\linewidth]{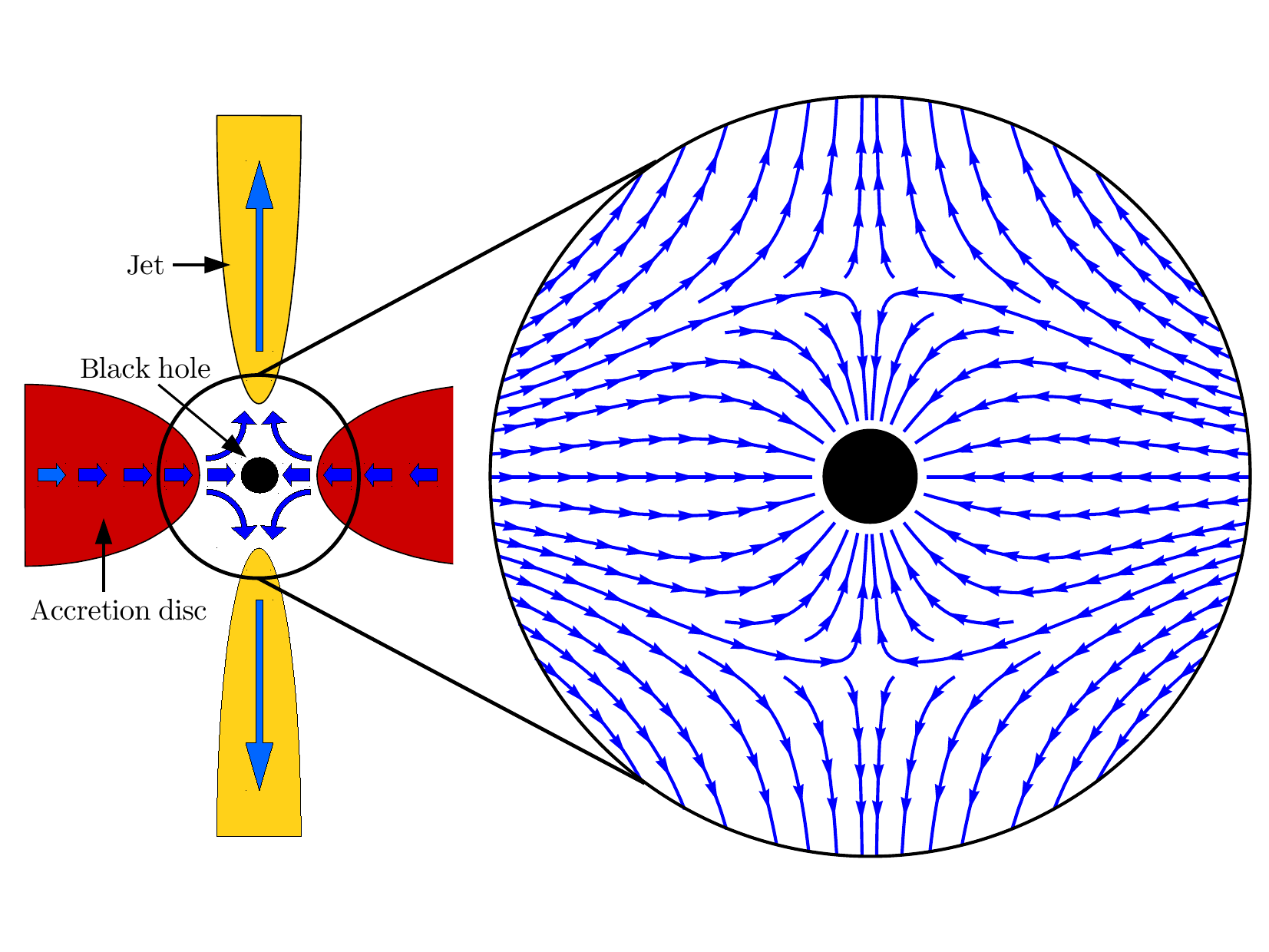} 
 \end{center}
\caption{Schematic representation of the astrophysical setting under study: the 
inner region of an accretion disk-jet system around a central black hole. The 
analytic solution presented in this work constitutes a toy model of the inner 
engine behind a jet-launching process in which, through the action of 
hydrodynamical forces only, an accretion flow can be transformed into an 
inflow-outflow bipolar structure. Subsequent numerical simulations relaxing the 
assumptions included in the analytic solution, validate and extend the 
qualitative aspects of the solution found to more general cases. The blue 
arrows show schematically the streamlines of the resulting flow. } 
\label{f1}
 \end{figure*}

Several mechanisms have been proposed to address these issues. The most widely 
accepted ones are the mechanisms introduced by \citet{BP1982} (BP) and 
\citet{BZ1977} (BZ). The BP mechanism consists on the extraction of energy and 
angular momentum from an accretion disk, via a magneto-centrifugal process. The 
main ingredient is a global, poloidal magnetic field threading an accretion 
disk that rotates with Keplerian velocity. This mechanism is mostly used to 
explain the origin of jets in AGNs~\citep[e.g.][]{hawley2015,bland2019} and in 
YSOs~\citep[e.g.][]{ouyed2003,pudritz2007,fendt2018}. On the other hand, the BZ 
mechanism shows an efficient way of extracting rotational energy from the spin 
of a Kerr black hole, provided a sufficiently strong magnetic field threads its 
event horizon.  This mechanism has been used to explain the jets associated to 
GRBs~\citep{lloyd2019,zhong2019} and radio jets in AGNs~\citep{Komissarov}.  
Both mechanisms have been successfully tested under broad physical conditions 
using general relativistic, magneto-hydrodynamic simulations 
\citep[e.g.][]{semenov04,McKinney06, qian2018,liska2019}.

On the other hand, a purely hydrodynamical mechanism has been proposed by 
\cite{hernandez14} in which an axisymmetric, polar density gradient is 
responsible for deflecting part of the material accreting from an equatorially 
over-dense inflow and redirecting it along a bipolar outflow. The main 
advantage of this jet-launching mechanism is that for it to work, one does not 
need to invoke the presence of magnetic fields that might lack the necessary 
strength or geometry in some systems \citep{hawley2015}, or processes taking 
place in the vicinity of a rotating event horizon and that, thus, can only 
account for jets associated with systems having a Kerr black hole as central 
accretor.

In this work, we revisit the jet-launching model of \cite{hernandez14} and 
study it in the general relativistic regime of an accreting, non-rotating black 
hole (Schwarzschild spacetime). Based on the general solution derived by 
\cite*{petrich88} for a relativistic potential flow with a stiff equation of 
state, we construct an analytic model corresponding to an inflow-outflow 
configuration around a Schwarzschild black hole. We propose that this analytic 
solution can be used as a toy model for the inner engine of a jet-launching 
system.

The physical setting of this model is shown schematically in Figure~\ref{f1}, 
and consists of the innermost region of an accretion disk-jet system around a 
central black hole. Specifically, we will confine our study to a finite, 
spherical region of radius $\mathcal{R}$ with the black hole at its center. We 
will refer to the surface of this domain as the injection sphere and consider 
it as the outer boundary of this system. Moreover,  for the analytic model 
presented, in addition to considering a perfect fluid described by a stiff 
equation of state, we will assume stationarity, axisymmetry, and an 
irrotational flow, i.e.~we consider that the gas entering the injection sphere 
from the inner edge of an accretion disk has lost all of its angular momentum 
through some kind of viscous dissipation mechanism 
\citep[e.g][]{shakura,balbus}. 

Even though for constructing the present model we do not include explicitly 
fluid rotation, it is important to remark that we have accounted for it 
indirectly by assuming that the flow configuration has a well defined symmetry 
axis, possibly as an inherited property of a rotation axis at larger scales. 
Furthermore, our assumption of a density anisotropy with the equatorial region 
having a higher density than the poles is a natural consequence of fluid 
rotation.

On the other hand, demanding a regular solution across the black hole event 
horizon implies that, for the present model with an ultrarelativistic stiff 
fluid, the total mass accretion rate onto the central black hole is fixed at a 
specific value \citep{petrich88}. This value corresponds closely to that found 
in the spherically {symme\-tric} case discussed by  \cite{michel72} for a 
Schwarzschild spacetime and by \cite{bondi52} in the non-relativistic regime. 

This important characteristic of the analytic model implies that the mass flux 
onto the central black hole is limited by a fixed value, and that any 
additional mass flux crossing the injection sphere has to be redirected and 
ejected from the system. In the present case, we show that the assumed 
anisotropic density field at the injection sphere translates into the bipolar 
outflow shown in Figure~\ref{f1}. Given that the incoming mass accretion rate 
is {\it choking} at a fixed value, we refer to this ejection mechanism as {\it 
choked accretion}.

With the aim of studying this accretion scenario under more general conditions, 
we also present the results of numerical simulations performed with the free 
GNU General Public License hydrodynamics code \textit{aztekas}\footnote{The code 
can be downloaded from \url{github.com/aztekas-code/aztekas-main}} 
\citep*{OM08,aguayo18,TA19}. By means of this numerical exploration, we are 
able to show that the choked accretion mechanism can operate for more realistic 
equations of state.

The basic idea behind the choked accretion model relies on a purely 
hydrodynamical mechanism and, thus, is not restricted to a relativistic regime. 
We presented the non-relativistic limit of the choked accretion model in  
\cite*{ATH19}. In that work we also introduced the Newtonian counterpart of the 
ultrarelativistic stiff fluid studied by \cite{petrich88} that, as discussed in 
\cite{tejeda18}, corresponds to the incompressible flow approximation.

With the present model we intend to draw attention to a potentially relevant 
phenomenon in which an accretion flow can become choked at a gravitational 
bottleneck, with the excess material being launched from the central region by 
a pure hydrodynamical mechanism. The present model is not intended as a 
substitute for other well-established jet-launching mechanisms, but rather as a 
further process based on simple physics that can operate alongside them.

The rest of this article is organized as follows. In Section~\ref{s2} we 
present the analytic toy model of choked accretion. In Section~\ref{s3} we 
explore numerically the feasibility of this model for fluids described by more 
realistic equations of state, where the constraint of potential flow imposed on 
the analytical model is dropped. There we find that the qualitative results of 
the analytic model also apply. We discuss possible astrophysical applications 
of 
the choked accretion model in Section~\ref{s4}.  Finally, in Section~\ref{s5} 
we summarize our results. Throughout this work we adopt geometrized units for 
which $\G = \cc = 1$. Greek indices denote spacetime components and we adopt 
the Einstein summation convention over repeated indices. 

\setcounter{equation}{0}
\section{Analytic model}
\label{s2}

In this section we present an analytic model of an inflow-outflow configuration 
around a Schwarzschild black hole. The model is based on the assumptions of a 
stationary, axisymmetry and irrotational flow. Moreover, we shall assume that 
the accreted gas corresponds to an ultrarelativistic gas described by a stiff 
equation of state of the form 
\begin{equation}
 P = K\rho^2, \label{e2.1}
\end{equation}
where $K=\co$, $P$ is the pressure, and $\rho$ is the rest mass 
density.\footnote{In relativistic hydrodynamics, it is customary to use the 
baryon number density $n$ instead of the rest mass density $\rho$. Introducing 
an average baryonic rest mass $m$, $n$ and $\rho$ are simply related as $\rho = 
m\,n$.}  

With the possible exception of the dense interior of a neutron star, the 
assumed stiff equation of state has a rather limited applicability in 
astrophysics \citep{lattimer07}. We have adopted this equation of state, 
however, as it allows us to carry out a full analytic treatment of the problem. 
The general relativistic solution obtained in this way, gives us a direct 
insight into the physics behind the proposed mechanism as well as the 
possibility to analyze in detail the dependence of the solution on the different 
model parameters. It is important to stress that this limiting assumption is 
relaxed in Section 3 where, by means of full-hydrodynamic simulations, we show 
that very similar results are obtained as steady state solutions for a more 
general equation of state and, thus, that the model here presented has a wider 
applicability in astrophysics.

For an ultrarelativistic gas one has that its internal energy $u$ is much 
larger than its rest mass energy, i.e.~$u\gg1$. This allows us to approximate 
the  corresponding specific enthalpy as $ h  = 1 + u + P/\rho \simeq u + 
P/\rho$. From the first law of thermodynamics together with the equation of 
state in \eq{e2.1},  it follows that $u = P/\rho$ and, hence
\begin{equation}
  h  = 2K\rho. \label{e2.2}
\end{equation}
From \eq{e2.2} it follows that, in the case of a stiff fluid, the sound speed 
$a$ is constant everywhere and equal to the speed of light, i.e.\footnote{Note 
that this definition of the sound speed is equivalent to the more common 
expression $a=\sqrt{\partial P/\partial e}$, where $e = \rho(1+u)$ is the 
relativistic energy density.}
\begin{equation}
 a \equiv \sqrt{\left.\left( \frac{\partial 
\ln h }{\partial\ln\rho}\right)\right|_s} = 1.
 \label{e2.3}
\end{equation}
This result implies that the corresponding flow will be subsonic at every point 
and that shock fronts can not develop.

\subsection{Potential flow}

The evolution of a perfect fluid in general relativity is dictated by local 
conservation equations, namely, the conservation of rest mass as expressed by 
the continuity equation
\begin{equation}
 (\rho\,U^\mu)_{;\mu} = 0 \label{e2.4}
\end{equation}
and local conservation of energy-momentum
\begin{equation}
 (T^\mu_\nu)_{;\mu} = \left( \rho\, h \,U^\mu\,U_\nu + P\,\delta^\mu_\nu 
\right)_{;\mu} = 0, \label{e2.5}
\end{equation}
where $U^\mu = \ud x^\mu / \ud \tau$ is the fluid four-velocity, $T^{\mu\nu}$ 
is 
the stress-energy tensor of a perfect fluid, $\delta^\mu_\nu$ is the Kronecker 
delta, and the semicolon stands for covariant differentiation. Since for a 
perfect fluid $\ud  h  = \ud P / \rho$, together with the continuity equation, 
\eq{e2.5} can be rewritten as
\begin{equation}
U^\mu( h \,U_\nu)_{;\mu} +  h _{,\nu} = 0. 
\label{e2.6}
\end{equation}

An irrotational flow is characterized by zero vorticity. In general 
relativity, vorticity is defined in terms of the tensor \citep{moncrief80}
\begin{equation}
 \omega_{\mu\nu} = 
P^\alpha_\mu\,P^\beta_\nu\left[\left( h \,U_\alpha\right)_{;\beta} - 
\left( h \,U_\beta\right)_{;\alpha}\right],
 \label{e2.7}
\end{equation}
where $P^\mu_\nu = U^\mu\,U_\nu + \delta^\mu_\nu$ is the projection tensor
onto the hypersurface orthogonal to $U^\mu$.

Expanding \eq{e2.7} and using \eq{e2.6} to simplify the resulting expression, 
we arrive at
\begin{equation}
 \omega_{\mu\nu} = \left( h \,U_\mu\right)_{;\nu} - 
\left( h \,U_\nu\right)_{;\mu}.
 \label{e2.8}
\end{equation}
From \eq{e2.8} we can see that a vanishing vorticity implies that 
$h \,U_\mu$ can be written as the gradient of a scalar velocity potential 
$\Phi$, i.e.
\begin{equation}
  h \,U_\mu = \Phi_{,\mu}.
 \label{e2.9}
\end{equation}

Substituting \eq{e2.9} into \eq{e2.4} leads to
\begin{equation}
 \left( \rho/ h \, \Phi^{,\mu} \right)_{;\mu} = 0.
\label{e2.10}
\end{equation}

In general we will have that $\rho$ is related to $ h $ through an equation 
of state while, from the normalization condition of $U^\mu$, $ h $ is related 
to $\Phi$ as $ h  = \sqrt{-\Phi_{,\mu}\Phi^{,\mu} }$. It is clear then that, 
in general, \eq{e2.10} will be a non-linear differential equation in $\Phi$ 
\citep[see e.g.][]{beskin95}. Nevertheless, by taking an ultrarelativistic 
fluid with a stiff equation of state (cf. Eq.~\ref{e2.2}), \eq{e2.10} reduces 
to the simple wave equation 
\begin{equation}
\Phi^{,\mu}_{\hspace{6pt};\mu} = 0.
\label{e2.11}
\end{equation}

In the case of Schwarzschild spacetime with spherical coordinates $(t, \,r, \, 
\theta, \,\phi)$, \eq{e2.11} has as general solution 
\citep{petrich88}
\begin{equation}
\Phi  = -e\,t  + \sum_{l,m} \left[A_{lm}\,P_l(\xi) + 
B_{lm}\,Q_l(\xi)\right] Y_{lm}(\theta,\,\phi) ,
\label{e2.12}
\end{equation}
where $Y_{lm}$ are spherical harmonics, $P_{lm}$, $Q_{lm}$ are Legendre 
functions on $\xi = r/M -1$, and $e$ is a constant related to the boundary 
conditions as we will show later on. \cite{petrich88} showed that requiring a 
regular solution across the black horizon necessarily implies that all $B_{lm}$ 
vanish identically except for $B_{00}$ which is in turn fixed as $B_{00} = 
4Me$. 

On the other hand, the coefficients $A_{lm}$ can be freely specified in order 
to match some given boundary conditions. In the present case, the assumption of 
axisymmetry leads us to consider only the $m=0$ modes, while demanding 
reflection symmetry with respect to the equatorial plane, leaves us only with 
even-$l$ multipoles different from zero. The lowest order model featuring both 
inflow and outflow regions can then be obtained from a velocity potential as in 
\eq{e2.12} with all $A_{lm} = 0$ except for $A_{20}$, i.e.
\begin{equation}
\begin{split}
\Phi  =\ & -e \bigg[ t  + 2M\ln\left(1-\frac{2M}{r}\right)  - \\
&\hspace{0.5cm} A \left(3\,r^2-6Mr+2M^2\right) \left(3\,\cos^2\theta -1\right) 
\bigg],
\end{split}
\label{e2.13}
\end{equation}
where $A = 4\sqrt{\pi/5} \,A_{20}/e$. 

Note that a different choice of the coefficients $A_{lm}$ will result in quite 
different flow configurations. For instance, \cite{petrich88,tejeda18} adopt 
the dipole $l=1$ to study the scenario of wind accretion.

With the velocity potential as given in \eq{e2.13} we have specified the 
dependence of the fluid properties on the polar angle $\theta$ at the outer 
boundary, i.e.~at the injection sphere $r=\mathcal{R}$. Nonetheless we are 
still free to specify the overall magnitude (scale) of the fluid properties at 
this boundary. In order to do this, we can specify values for the fluid 
velocity, density, and pressure (or any other pair of thermodynamical variables) 
at a reference point on the injection sphere. For this work, we shall take as 
reference the point $(r=\mathcal{R},\,\theta = \pi/2)$, i.e.~the equator of the 
injection sphere. Let us call $\rho_0$ and $P_0$ the values of the density and 
pressure at this point as measured by a co-moving observer. Clearly, from these 
reference values we can write $K = P_0 / \rho_0^2$ and $h_0 = 2\,P_0/\rho_0$. 
On the other hand, we parametrize the fluid velocity at this point using $V_0$, 
defined as the magnitude of the three-velocity vector measured by a local 
Eulerian observer (LEO).\footnote{These are static observers carrying a local 
tetrad with respect to which they can perform local measurements, thus 
describing physical properties of the fluid. This family of observers can be 
introduced in a covariant (coordinate-independent) way by noticing that their 
four-velocity corresponds to the time isometry of Schwarzschild spacetime as 
encoded by the time-like Killing vector $t^\mu = \delta^\mu_t$.} In terms of 
$V_0$, the four-velocity of the fluid at the equator of the injection sphere is 
given by
\begin{equation}
U^\mu = \Gamma_0\left( 1/\alpha_0,\ -\alpha_0\,V_0,\ 0,\ 0\right), 
\label{e2.14}
\end{equation}
with $\alpha_0 = \alpha(\mathcal{R})$, where
\begin{equation}
\alpha = \sqrt{1-\frac{2M}{r}} 
\label{e2.15}
\end{equation}
is the lapse function associated to the 3+1 decomposition of the four-metric and
\begin{equation}
 \Gamma_0 = \left(1 - V_0^2\right)^{-1/2}
 \label{e2.16}
\end{equation}
is the Lorentz factor between the fluid element and the LEO. As $V_0$ 
corresponds to the magnitude of a physical three-velocity vector, it is 
naturally bounded as $V_0 < 1$. Also note that in \eq{e2.14} we have explicitly 
considered that the radial velocity is negative at the reference point  as we 
are interested in a scenario with equatorial inflow. The velocity potential in 
\eq{e2.13} should also be useful to describe a very different scenario with 
polar inflow and equatorial outflow (akin to a wall jet) by allowing for a 
positive radial velocity at the reference point. We shall only focus on the 
former case for the remainder of this work.

It is worth noticing at this point that the present analytic model is scale 
free with respect to the specific values of $M$, $\rho_0$ and $P_0$. On the 
other hand, as we shall see below, the parameters dictating the overall 
morphology of the resulting accretion flow are $V_0$ and $\mathcal{R}/M$.

\subsection{Velocity field}

Substituting the velocity potential $\Phi$ given in \eq{e2.13} into \eq{e2.9} 
leads to the velocity field
\begin{align}
&\frac{h}{e}\frac{\ud t}{\ud \tau}  = 
 \left(1-\frac{2M}{r}\right)^{-1} ,
\label{e2.17}\\
&\frac{h}{e}\frac{\ud r}{\ud \tau}   =  - \frac{4M^2}{r^2} +  
\frac{6A}{r}(r-M)(r-2M)\left(3\cos^2\theta -1\right) ,
\label{e2.18} \\
& \frac{h}{e}\frac{\ud \theta}{\ud \tau}  = 
-\,\frac{6A}{r^2}
\left(3\,r^2-6Mr+2M^2\right)\sin\theta\cos\theta. 
\label{e2.19}
\end{align}

By evaluating \eqs{e2.17} and \eqref{e2.18} at the reference point 
$(r=\mathcal{R},\,\theta = \pi/2)$ and comparing the result with \eq{e2.14}, we 
arrive at the following expressions for the constants $e$ and $A$ in terms of 
the boundary conditions
\begin{gather}
 e = \alpha_0\,h_0\,\Gamma_0, \label{e2.20}\\
 A = \frac{V_0 \mathcal{R}^2 - 
4M^2}{6\mathcal{R}(\mathcal{R}-M)(\mathcal{R}-2M)}.
 \label{e2.21}
\end{gather}

Note that, since we have assumed inflow across the equatorial region, the 
velocity field described by \eqs{e2.18} and \eqref{e2.19} is characterized by 
the existence of a pair of stagnation points (points at which the spatial 
components of the velocity field vanish) located along the polar axis ($\theta 
= 0,\,\pi$) at mirror points with respect to the origin. Calling $\mathcal{S}$ 
their radial distance to the origin, from \eq{e2.18} we obtain the following 
relationship between $V_0$, $\mathcal{R}$ and $\mathcal{S}$
\begin{equation}
 V_0 = \frac{2M^2}{\mathcal{R}^2}\left[2 + 
 \frac{\mathcal{R}(\mathcal{R}-M)(\mathcal{R}-2M)}
 {\mathcal{S}(\mathcal{S}-M)(\mathcal{S}-2M)} \right].
 \label{e2.22}
\end{equation}
 Alternatively, \eq{e2.22} can be inverted to express $\mathcal{S}$ as a
function of $V_0$ and $\mathcal{R}$
\begin{equation}
 \frac{\mathcal{S}}{M} = 1  + \left( \xi + \sqrt{\xi^2-\frac{1}{27}} 
\right)^{1/3} 
   + \left( \xi - \sqrt{\xi^2-\frac{1}{27}} \right)^{1/3} ,
\label{e2.23}
\end{equation}
where
\begin{equation}
 \xi = 
\frac{\mathcal{R}(\mathcal{R}-M)(\mathcal{R}-2M)}{M(V_0\mathcal{R}^2-4M^2)}.
\label{e2.24} 
\end{equation}

Note that we can also use $\mathcal{S}$ to rewrite the coefficient $A$ as
\begin{equation}
 A = \frac{M^2}{3\,\mathcal{S}(\mathcal{S}-M)(\mathcal{S}-2M)}.
 \label{e2.25}
\end{equation}

Using \eq{e2.25}, together with \eq{e2.17} to get rid of the dependence on $h$, 
we can rewrite the spatial components of the velocity with respect to the 
coordinate time $t$ as
\begin{align}
\frac{\ud r}{\ud t} & =  -
\frac{2M^2}{r^2}\left(1-\frac{2M}{r} \right) \times \nonumber \\
& \hspace{0.5cm}\left[ 2 -
\frac{r(r-M)(r-2M)}{\mathcal{S}(\mathcal{S}-M)(\mathcal{S}-2M)}
\left(3\cos^2\theta -1\right)
\right], \label{e2.26} \\
\frac{\ud \theta}{\ud t} & = 
-\frac{2M^2}{r^2}\left(1-\frac{2M}{r} \right)
\frac{3\,r^2-6Mr+2M^2}{\mathcal{S}(\mathcal{S}-M)(\mathcal{S}-2M)}
\sin\theta\cos\theta. 
\label{e2.27}
\end{align}

Furthermore, we can also express the velocity field in terms of the physical, 
locally measured components of the three-velocity defined by LEOs and given by
\begin{gather}
 V^r = \left(1-\frac{2M}{r}\right)^{-1/2}\frac{\ud r}{\ud t},
 \label{e2.28}\\
 V^\theta = \left(1-\frac{2M}{r}\right)^{-1/2}\frac{\ud \theta}{\ud t},
 \label{e2.29}
\end{gather}
as well as its corresponding (squared) magnitude 
\begin{equation}
\begin{split}
 V^2 &  = \left(1-\frac{2M}{r}\right)^{-2}\left(\frac{\ud r}{\ud t}\right)^2 + 
\left(1-\frac{2M}{r}\right)^{-1}r^2\left(\frac{\ud \theta}{\ud t}\right)^2 \\
 & = \frac{4M^4}{r^4}\Bigg[ 4 - 
\frac{4\,r(r-M)(r-2M)}{\mathcal{S}(\mathcal{S}-M)(\mathcal{S}-2M)}
(3\cos^2\theta -1) + \\
 & \hspace{0.5cm} 
\frac{r^2(r-M)^2(r-2M)^2}{\mathcal{S}^2(\mathcal{S}-M)^2(\mathcal{S}-2M)^2} 
(3\cos^2\theta -1)^2 +  \\
 & \hspace{0.5cm}  
\frac{r(r-2M)\left(3\,r^2-6Mr+2M^2\right)^2}{\mathcal{S}^2(\mathcal{S}
-M)^2(\mathcal{S}-2M)^2}  \sin^2\theta\,\cos^2\theta  \Bigg].
\end{split} 
 \label{e2.30}
\end{equation}

Note that for a sufficiently large radius $r$ the physical three-velocity 
magnitude $V$ as given in \eq{e2.30} will grow like $V\propto r$, eventually 
becoming superluminal.\footnote{In terms of the velocity potential $\Phi$ this 
translates into the gradient $\Phi_{,\mu}$ transitioning from being timelike to 
spacelike.} To prevent this from happening, we need to consider the preset 
model as a local solution that is only properly defined within a finite spatial 
domain. For simplicity we will restrict this work to the spherical domain $r\le 
\mathcal{R}$, where $\mathcal{R}$ is the radius of the injection sphere. 

By examining \eq{e2.30} we can see that, for a radius $ \mathcal{R}> 
\mathcal{S}$, $V$ reaches its maximum at the polar axis, which is given by
\begin{align}
V(\mathcal{R},\ 0) & = \frac{4M^2}{\mathcal{R}^2}\left| 
\frac{\mathcal{R}(\mathcal{R}-M)(\mathcal{R}-2M)}{\mathcal{S}(\mathcal{S}
-M)(\mathcal{S}-2M)}-1 \right| \nonumber \\
 & = \left|2\,V_0 - \frac{12 M^2}{\mathcal{R}^2}\right| ,
  \label{e2.31}
\end{align}
where we have used \eq{e2.22} to arrive at the last equal sign. From this last 
expression, we  obtain the following upper bound on $V_0$ in order to guarantee 
$V$ to be subluminal within the domain $2M < r\le \mathcal{R}$\footnote{ 
Note that at the event horizon  $V(2M) = 1$, although this is only due to 
the fact that the Eulerian observers become ill-defined at this radius. The 
fluid velocity as described by $U^\mu$ is completely regular across the 
horizon.}
\begin{equation}
 V_0 < \frac{1}{2} + 6\frac{M^2}{\mathcal{R}^2}.
 \label{e2.32}
\end{equation}

Based on \eq{e2.31}, and taking into account the sign of the radial velocity in 
\eq{e2.26}, we define the ejection velocity at the poles of the shell 
$r=\mathcal{R}$ as
\begin{equation}
 V_\mathrm{ej} \equiv 2\,V_0 - \frac{12 M^2}{\mathcal{R}^2}.
 \label{e2.33}
\end{equation}
From this expression we note that, in order to actually have polar outflow 
at $r=\mathcal{R}$ (i.e. $V_\mathrm{ej}>0$), we require
\begin{equation}
 V_0 > 6\frac{M^2}{\mathcal{R}^2}.
 \label{e2.34}
\end{equation}
Moreover, note that when $0<V_0< 6\frac{M^2}{\mathcal{R}^2}$ the stagnation 
point lies outside the injection sphere ($\mathcal{S}>\mathcal{R}$) and the 
flow is everywhere radially inwards although not spherically symmetric.

Summarizing the previous results, only for values of $V_0$ within the range
\begin{equation}
6\frac{M^2}{\mathcal{R}^2} < V_0 < \frac{1}{2} + 6\frac{M^2}{\mathcal{R}^2}
\label{e2.35}
\end{equation}
we find flow configurations characterized by equatorial inflow and bipolar 
outflows within the domain of interest $2M < r\le\mathcal{R}$.

 \begin{figure*}
 \begin{center}
  \includegraphics[scale=0.5]{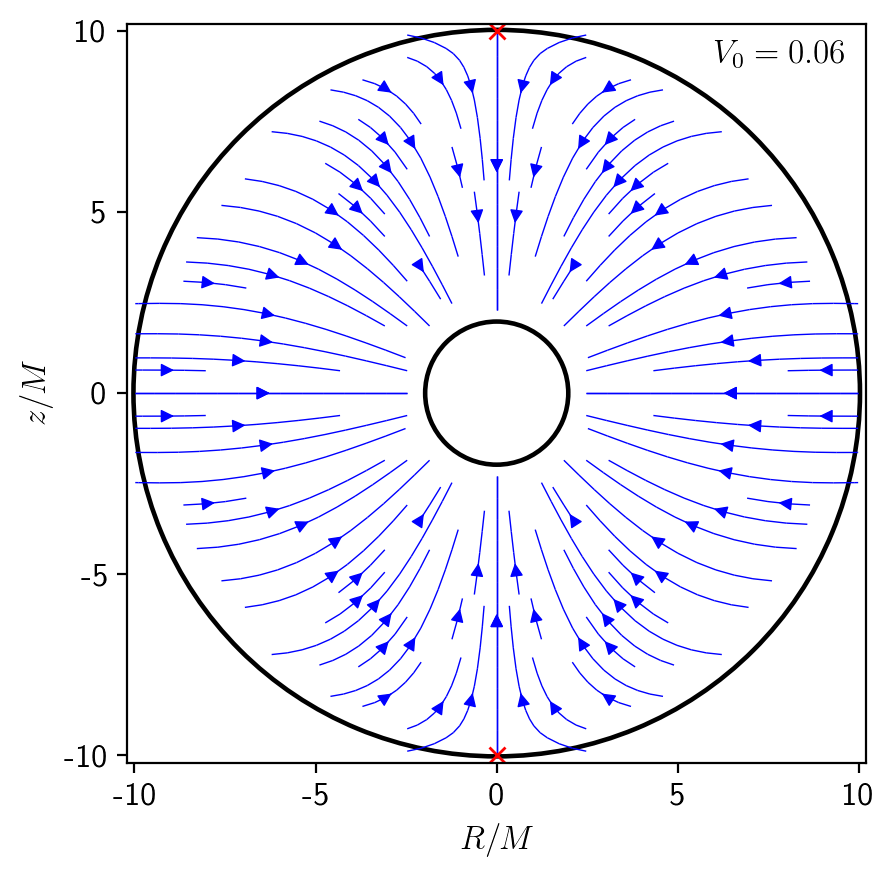}  
  \includegraphics[scale=0.5]{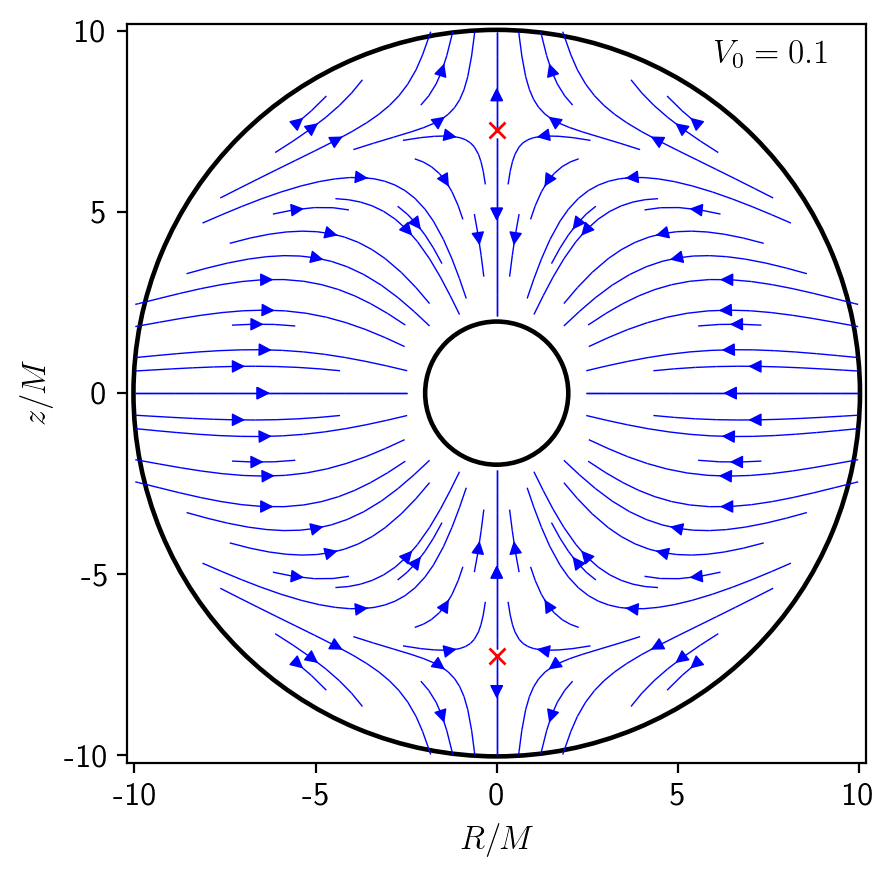}
  \includegraphics[scale=0.5]{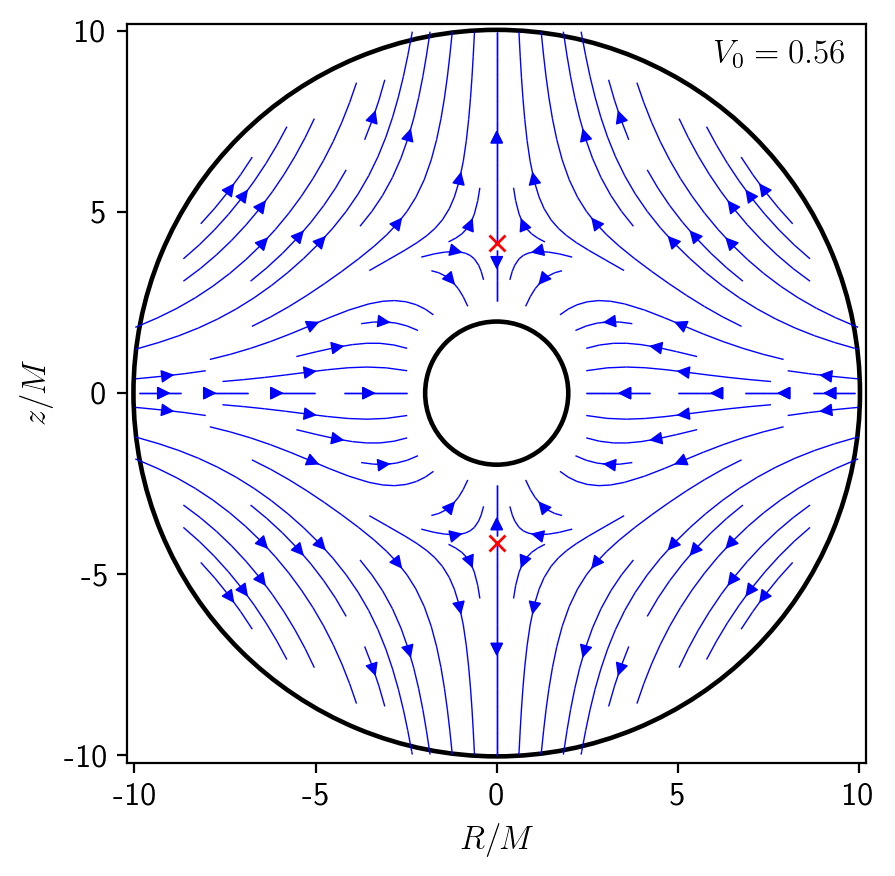}
  \end{center}
\caption{  Streamlines of the accretion flow resulting from the velocity field 
in \eqs{e2.26} and \eqref{e2.27}. We have taken $\mathcal{R} = 10M$ as radius 
of the injection sphere while, from left to right, $V_0 = 0.06,\,0.1,\,0.56$. 
Note that the first and third values of $V_0$ correspond to the lower and upper 
limits in \eq{e2.35}, respectively. The stagnation points in each case are 
shown as red crosses. The outer boundary of the model $(r=\mathcal{R})$ as well 
as the event horizon of the central black hole $(r=2M)$ are shown as circles 
drawn with thick, solid lines. The axes correspond to the usual cylindrical 
coordinates $R = r\,\sin\theta$, $z = r\,\cos\theta$. }
\label{f2}
 \end{figure*}%

See Figure~\ref{f2} for three examples of the streamlines resulting from the 
velocity field in \eqs{e2.26} and \eqref{e2.27} for $\mathcal{R} = 10M$. On the 
left and right panels $V_0 = 0.06,\,0.56$, which correspond to the lower and 
upper bounds of the interval in \eq{e2.35}. For the central panel we have taken 
$V_0 = 0.1$ as a representative middle value for $V_0$. 

\begin{figure}
 \begin{center}
  \includegraphics[trim=0 39 0 0, clip, width=\linewidth]{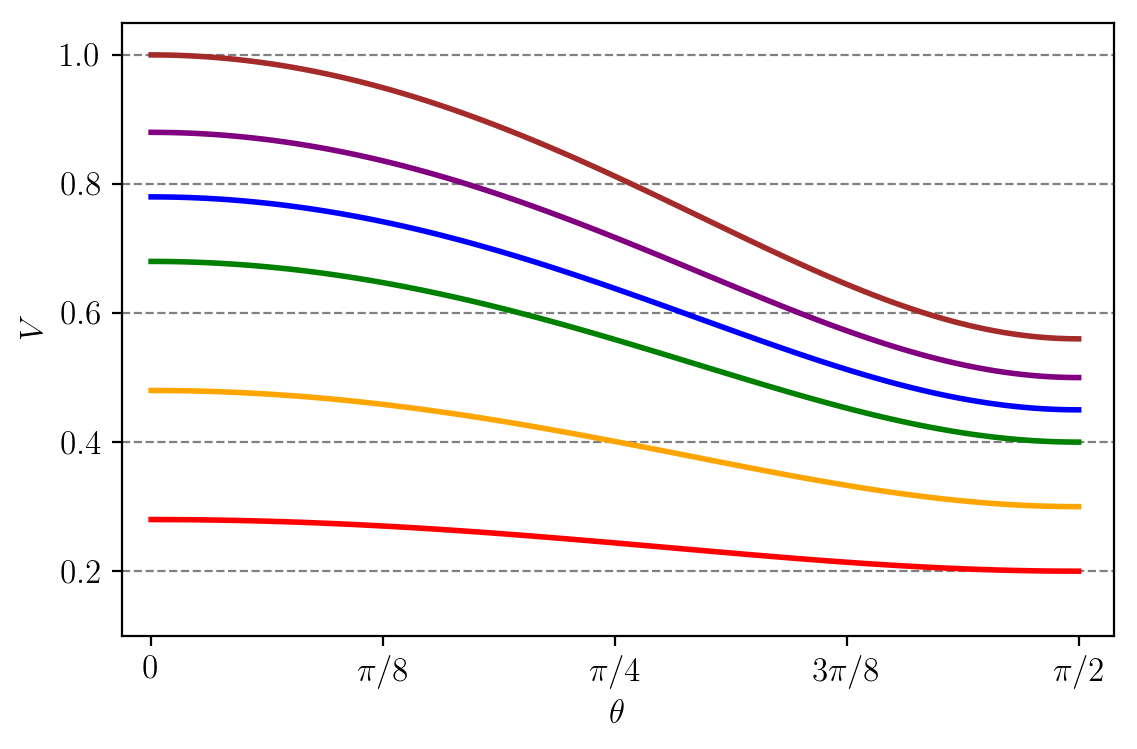}  
  \includegraphics[width=\linewidth]{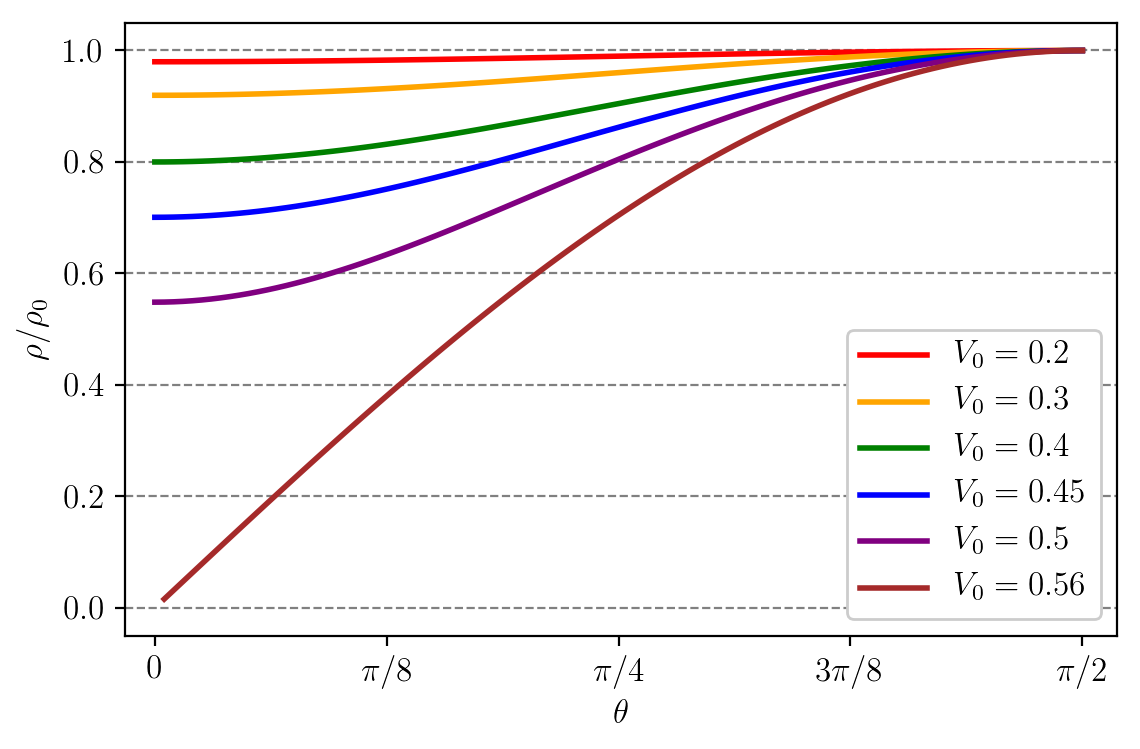}
  \end{center}
\caption{Magnitude of the three-velocity $V$ (Eq.~\ref{e2.30}) and density 
$\rho$ (Eq.~\ref{e2.36}) of the analytic model for an ultrarelativistic stiff 
fluid.  Both quantities are shown as functions of the polar angle $\theta$ 
evaluated at the injection sphere for the particular case $\mathcal{R} = 10M$ 
and six different values of the velocity $V_0$. In this case, from \eq{e2.32} 
we have that $V_0$ is limited as $V_0 < 0.56$ in order to guarantee that  the 
whole solution is well defined within the spatial domain $r\le \mathcal{R}$. 
Note that for $V_0=0.56$, $V_\mathrm{ej} =1 $ while the corresponding value of 
$\rho$ goes to zero.}
\label{f3}
 \end{figure}

In the top panel of Figure~\ref{f3} we show the magnitude of the three-velocity 
$V$ as function of the polar angle $\theta$ evaluated at the injection sphere 
for the particular case $\mathcal{R}=10M$ and several values of $V_0$. 

 \begin{figure*}
 \begin{center}
  \includegraphics[scale=0.5]{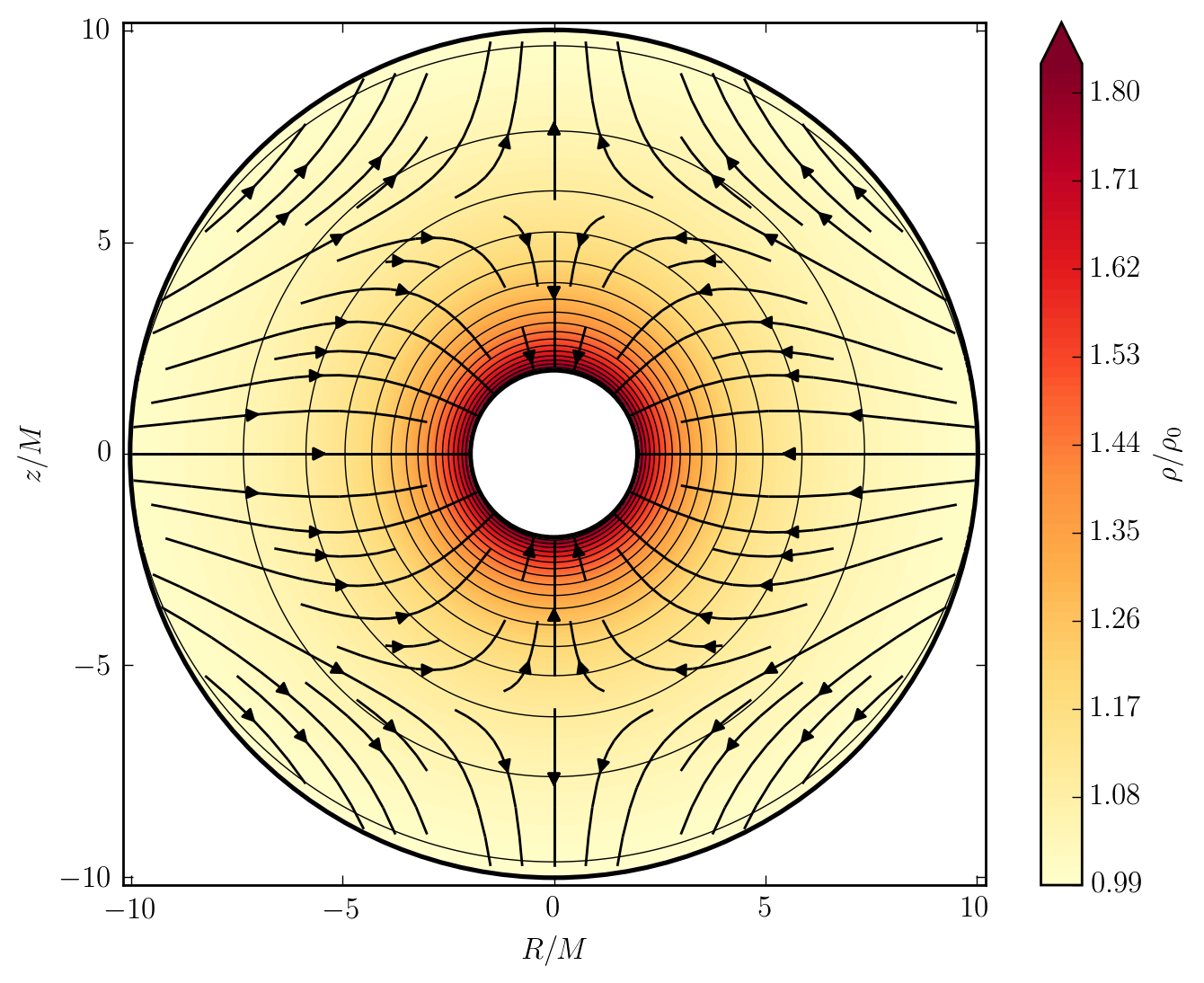} 
  \includegraphics[scale=0.5]{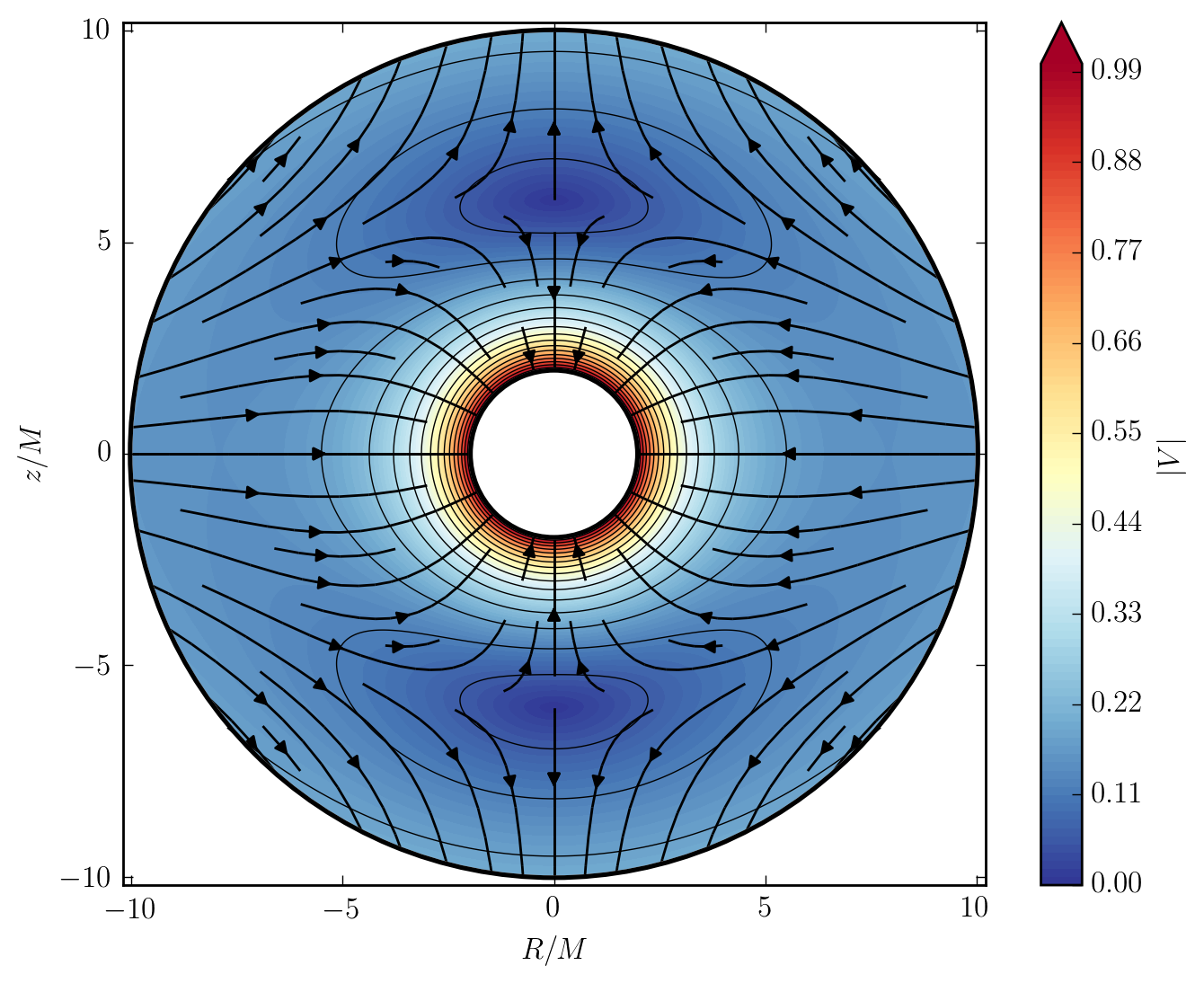} 
 \end{center}
\caption{Example of the analytic model of choked accretion for the values 
$\mathcal{R}= 10M$ and $V_0 = 0.16$. The figure shows isocontours of the 
fluid's density as given by \eq{e2.36} (left panel) as well as the magnitude of 
the three-velocity as given by \eq{e2.30} (right panel). Note that the 
stagnation points are located at $\mathcal{S}= 6M$. Fluid streamlines are 
indicated by thick, solid lines with an arrow. The axes correspond to the usual 
cylindrical coordinates $R = r\,\sin\theta$, $z = r\,\cos\theta$.}
\label{f4}
 \end{figure*}

\subsection{Density field}

We can now recover the density field by substituting \eqs{e2.17}--\eqref{e2.19} 
into the normalization condition of the four-velocity $U_\mu U^\mu = -1$ and 
then using \eq{e2.20}; the result is
\begin{equation}
 \frac{\rho}{\rho_0}  = \frac{\alpha_0\,\Gamma_0}{\alpha\,\Gamma} =
 \sqrt{ \frac{r(\mathcal{R}-2M)(1-V^2)}{\mathcal{R}(r-2M)(1-V_0^2)} },
\label{e2.36}
\end{equation}
with $V^2(r,\theta)$ as given in \eq{e2.30}. 

Recalling that the local density measured by a LEO is given by $D = \Gamma 
\rho$, from \eq{e2.36} we obtain the interesting result that the density field 
as described by LEOs is spherically symmetric, i.e.~$D$ is only a function of 
$r$.

Note that the same criterion introduced in \eq{e2.32} in order to guarantee a 
subluminal three-velocity within $2M < r\le \mathcal{R}$, also guarantees that 
the density field as expressed in \eq{e2.36} is a well-defined, real quantity 
within the same spatial domain. 

From \eq{e2.36} we obtain the following simple relation for the ratio between 
the density at the pole and the equator of the injection sphere
\begin{equation}
\begin{split}
 \frac{\rho(\mathcal{R},0)}{\rho(\mathcal{R},\pi/2)} & = 
\frac{\rho(\mathcal{R},0)}{\rho_0} = 
 \sqrt{ \frac{1-V^2_\mathrm{ej}}{1-V_0^2} }  \\
 & = \sqrt{ \frac{\mathcal{R}^4 - 4\,V_0^2\mathcal{R}^4 + 48 V_0\mathcal{R}^2 
M^2 - 144M^4}{\mathcal{R}^4-V_0^2\mathcal{R}^4} } .
 \label{e2.37}
\end{split}
\end{equation}

In the following we shall use the contrast $\delta$ between the polar and 
equatorial densities at the injection sphere defined as  
\begin{equation}
 \delta = 1 - \frac{\rho(\mathcal{R},0)}{\rho_0}.
\label{e2.38}
\end{equation}
From \eqs{e2.37} and \eqref{e2.38} we can see that an arbitrarily small density 
contrast suffices not only to produce the inflow-outflow configuration shown in 
the central and right panels of Figure~\ref{f2}, but also to guarantee that 
$V_\mathrm{ej} > V_0$. Furthermore notice that, as the density contrast 
approaches unity, the ejection velocity approaches the speed of light. Indeed, 
for the present case of an ultrarelativistic stiff fluid, as 
$\delta\rightarrow1$ we can obtain arbitrarily large Lorentz factors for the 
ejected flow.

Complementary to the top panel of Figure~\ref{f3}, where we see that the 
magnitude of the velocity field at the injection sphere increases as $V_0$ 
increases, in the bottom panel of this figure we show the angular density 
profile $\rho(\theta)$ evaluated at the injection sphere. From this figure we 
see that, as $V_0$ increases, the polar to equatorial density contrast 
increases. Moreover, we can also see that as the velocity at the poles becomes 
luminal for $V_0 = 1/2 + 6M^2/\mathcal{R}^2$, the corresponding value of the 
density field becomes zero. 

\subsection{Equation for the streamlines}

An equation for the streamlines can be found by combining \eqs{e2.26} and 
\eqref{e2.27} to obtain
\begin{equation}
 \frac{\ud r}{\ud \theta} = \frac{ 2 -
\frac{r(r-M)(r-2M)}{\mathcal{S}(\mathcal{S}-M)(\mathcal{S}-2M)}
\left(3\cos^2\theta -1\right) }{
\frac{3\,r^2-6Mr+2M^2}{\mathcal{S}(\mathcal{S}-M)(\mathcal{S}-2M)}
\sin\theta\cos\theta}, 
\label{e2.38**}
\end{equation}
which in turn can be integrated as
\begin{equation}
 \Psi = 
\cos\theta\left[1+\frac{r(r-M)(r-2M)}{\mathcal{S}(\mathcal{S}-M)(\mathcal{S}-2M)
}\frac{\sin^2\theta}{2} \right],
 \label{e2.40}
\end{equation}
where $\Psi$ is an integration constant. \eq{e2.40} constitutes an implicit 
equation for the streamlines, where, for every constant value of $\Psi$, one 
has a different streamline. Note in particular that $\Psi = \pm 1$ corresponds 
to the streamlines reaching the stagnation points located at 
$(r=\mathcal{S},\,\theta = 0)$ for the plus sign and $(r=\mathcal{S},\,\theta = 
\pi)$ for the minus sign. Streamlines with $|\Psi|<1$ end up accreting onto the 
central black hole, while those with $|\Psi|>1$ escape along the bipolar 
outflow.

In Figure~\ref{f4} we show the resulting density, velocity and streamlines of 
the analytic model of choked accretion for the particular values of 
$\mathcal{R}= 10M$, $V_0 = 0.16$.  For this choice of boundary conditions
the stagnation points are located at $\mathcal{S}= 6M$.

\subsection{Mass accretion, injection and ejection rates}

The total mass accretion rate onto the central black hole can be calculated as 
the flux of mass density integrated over any closed surface $\sigma$ 
enclosing it, i.e.
\begin{equation}
 \dot{M}  =  - \int_\sigma \rho\,U^\mu\,\sqrt{-g}\,\ud S_\mu ,
 \label{e2.41}
\end{equation}
where $\sqrt{-g} = r^2\sin\theta$ and $\ud S_\mu$ is a differential area 
element orthogonal to the surface $\sigma$. Taking any sphere of radius $r$ as 
the integration surface, together with the conditions of axisymmetry and 
stationarity, we obtain 
\begin{equation}
\begin{split}
 \dot{M}  & =  - 2\pi \int_0^\pi \rho\,U^r\,r^2\sin\theta\,\ud\theta \\
  & = 16\pi M^2\alpha_0\rho_0\Gamma_0.
\end{split}
\label{e2.42}
\end{equation} 
The result of \eq{e2.42} holds even if higher multipoles are considered in the 
velocity potential (cf.~Eq.~\ref{e2.12}): by virtue of the orthogonality of the 
spherical harmonics, the contribution of any multipole $(l,m)$ to the integral 
in \eq{e2.42} identically vanishes except for the spherically symmetric 
monopole $l=0,\,m=0$. Note however that, in the spherically symmetric case, 
$V_0$ is not a free parameter. In accordance with \eq{e2.30}, in this case $V_0 
= 4M^2/\mathcal{R}^2$. Therefore, in the spherically symmetric case, the mass 
accretion rate as given by \eq{e2.42} can be written as
\begin{equation}
 \dot{M}_\mathrm{M} = 16\pi M^2\alpha_0\rho_0\left(1 
-16\frac{M^4}{\mathcal{R}^4}\right)^{-1/2}.
 \label{e2.43}
\end{equation}
This value corresponds to the \cite{michel72} solution as applied to a stiff 
equation of state, as shown by \cite{chaverra15}. See Appendix \ref{sA} for a 
brief overview of the \cite{michel72} model in the case of a general polytrope.

We can express the general result for the mass accretion rate as given in 
\eq{e2.42} in units of $\dot{M}_\mathrm{M}$ as
\begin{equation}
 \dot{M} = \eta\, \dot{M}_\mathrm{M},
 \label{e2.44}
\end{equation}
where
\begin{equation}
 \eta = \left( \frac{1 - 16M^4/\mathcal{R}^4}{1 - V_0^2 } \right)^{1/2}.
 \label{e2.45}
\end{equation}
Note that in most cases of interest $\eta\gtrsim 1$. For instance, taking 
$\mathcal{R} = 10 M$, from the allowed range of velocities in \eq{e2.35}, 
we obtain $1 < \eta \lesssim 1.2$, while for $\mathcal{R} \gg M$, 
we have $1 < \eta \lesssim 1.15$.

On the other hand, we can also define the mass injection rate 
$\dot{M}_\mathrm{in}$ as the inward flux of mass across the injection sphere of 
radius $\mathcal{R}$, i.e.~by considering an integration analogous to the one 
in \eq{e2.42} but in which we consider only the fluid elements with a negative 
radial velocity $U^r$. From \eq{e2.18} we obtain that  $U^r(\mathcal{R},\theta) 
< 0$ for $\theta_c < \theta < \pi - \theta_c$, where $\theta_c$ is such that 
$U^r(\mathcal{R},\theta_c) = 0$  and is given by
\begin{equation}
\begin{split}
 \cos\theta_c & = 
\sqrt{\frac{1}{3}+\frac{2}{3}\,\frac{\mathcal{S}(\mathcal{S}-M)(\mathcal{S}-2M)}
{\mathcal{R}(\mathcal{R}-M)(\mathcal{R}-2M)}} \\
 & = \sqrt{\frac{V_0\mathcal{R}^2}{3(V_0\mathcal{R}^2 - 4M^2)}}.
\end{split}
\label{e2.46}
\end{equation}

We can thus calculate $\dot{M}_\mathrm{in}$ as 
\begin{equation}
 \dot{M}_\mathrm{in}  =
 - 4\pi \int_{\theta_c}^{\pi/2} \rho\,U^r\,r^2\sin\theta\,\ud \theta  =  
 \Lambda\, \dot{M} ,
 \label{e2.47}
\end{equation}
with 
\begin{equation}
\begin{split}
 \Lambda & = \frac{1}{6\sqrt{3}} \frac{V_0\mathcal{R}^2}{M^2} \left( 1 - 
\frac{4M^2}{V_0\mathcal{R}^2} \right)^{-1/2} \\
  & \simeq \frac{1}{6\sqrt{3}}\left(  \frac{V_0\mathcal{R}^2}{M^2} + 2 \right)
  + \mathcal{O}\left(\frac{M^2}{V_0\mathcal{R}^2} \right) ,
\end{split}
 \label{e2.48}
\end{equation} 
where for the second step we have used the Taylor series expansion 
assuming $M^2 \ll V_0 \mathcal{R}^2$. 

From \eq{e2.46} we have that when $V_0 = 6M^2/\mathcal{R}^2$ (or, equivalently 
$\mathcal{S}=\mathcal{R}$) then $\theta_c = 0$ and hence $\dot{M}_\mathrm{in} = 
\dot{M}$. On the other hand, when $V_0 < 6M^2/\mathcal{R}^2$ we have that 
$U^r<0$ for all $\theta$ and again $\dot{M}_\mathrm{in} = \dot{M}$. We can then 
write
\begin{equation}
    \dot{M}_\mathrm{in}  = 
  \begin{cases}
      \dot{M} , & \text{if}\ V_0 \le 6M^2/\mathcal{R}^2 , \\
      \Lambda\,\dot{M}, & \text{if}\ V_0 > 6M^2/\mathcal{R}^2 .
  \end{cases}
\label{e2.49}  
\end{equation}

Similarly, we define the mass ejection rate $\dot{M}_\mathrm{ej}$ as the 
outward flux of mass across the sphere of radius $r = \mathcal{R}$. Clearly 
\begin{equation}
 \dot{M}_\mathrm{ej} = 
  \begin{cases}
      0, & \text{if}\  V_0 \le 6M^2/\mathcal{R}^2 , \\
      (\Lambda - 1)\dot{M} , & \text{if}\ V_0 > 6M^2/\mathcal{R}^2 .
  \end{cases}
\label{e2.50}  
\end{equation}

Note that for a fixed injection radius $\mathcal{R}$, it can be shown that the 
upper bound for $V_0$ found in \eq{e2.32} implies the following upper bound for 
the mass injection rate
\begin{equation}
 \dot{M}_\mathrm{in} < 8\pi\,\rho_0 
\sqrt{\frac{(\mathcal{R}-2M)\mathcal{R}^3(\mathcal{R}^2+12M^2)^3}{3(\mathcal{R}
^2 + 4M^2)
[4\mathcal{R}^4 - (\mathcal{R}^2+12M^2)^2]}}.
\label{e2.51} 
\end{equation}

The \eqs{e2.49} and \eqref{e2.50} encapsulate the concept of choked accretion 
described in the introduction:  the central black hole accretes at an 
essentially fixed  rate $\dot{M} \simeq \dot{M}_\mathrm{M}$. Whenever the mass 
injection rate surpasses this limit, the excess flux is ejected from the system 
as a bipolar outflow at the rate $\dot{M}_\mathrm{ej}$.  Note in particular 
that from \eq{e2.50} we can write
\begin{equation}
 \frac{\dot{M}_\mathrm{ej}}{\dot{M}_\mathrm{in}} = 1 - 
\frac{\dot{M}}{\dot{M}_\mathrm{in}}  = 1 - \frac{1}{\Lambda}.
 \label{e2.52} 
\end{equation}
This simple functional dependence of the ratio of ejected to injected mass 
fluxes on the injection mass rate is shown in Figure~\ref{f14} to compare the 
analytic model against the results of hydrodynamic numerical simulations.

 \begin{figure}
 \begin{center}
  \includegraphics[trim=0 42 0 0, clip,width=\linewidth]{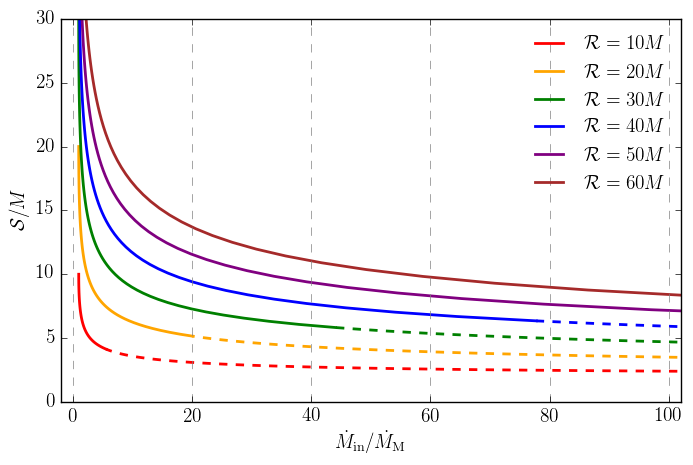}  
  \includegraphics[trim=0 42 0 0, clip,width=\linewidth]{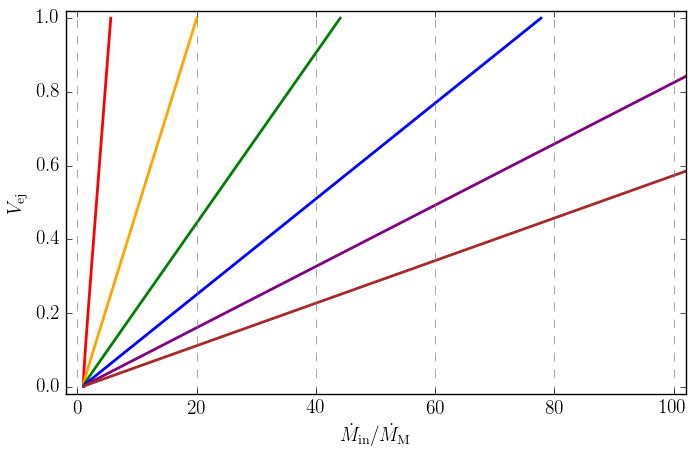}
  \includegraphics[width=\linewidth]{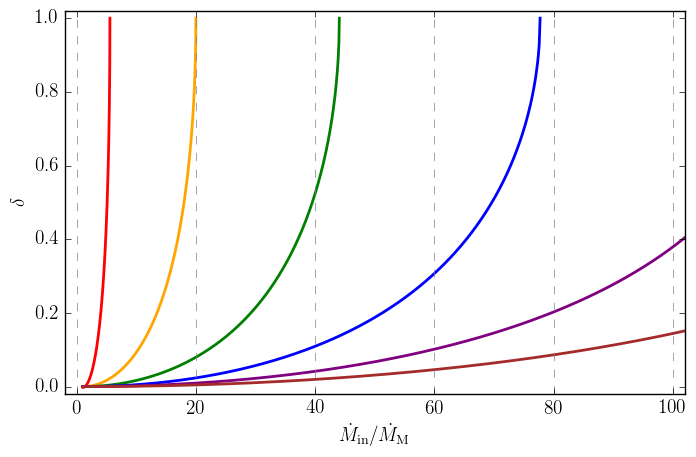}
  \end{center}
\caption{ Dependence of different properties of the analytic model of choked 
accretion on the parameters $\mathcal{R}/M$ and 
$\dot{M}_\mathrm{in}/\dot{M}_\mathrm{M}$. From top to bottom, each panel  
shows: the location of the stagnation point $\mathcal{S}$, the maximum velocity 
attained by the ejected material $V_\mathrm{ej}$ (Eq.~\ref{e2.31}), and the 
density contrast between the pole and the equator of the injection sphere 
$\delta$ (Eq.~\ref{e2.38}). The dashed lines in the top panel indicate regions 
in the parameter space for which the model is not well defined within the whole 
domain $r<\mathcal{R}$ (see discussion in the main text).}
\label{f5}
 \end{figure}%

Finally, it is interesting to explore the behavior of the analytic model as a 
function of the parameters $\mathcal{R}$ and 
$\dot{M}_\mathrm{in}/\dot{M}_\mathrm{M} = \eta\,\Lambda$. Note that, as shown 
in \eq{e2.48}, $\eta\,\Lambda$ is essentially a linear reparametrization of 
$V_0$ for most of the domain of interest. In Figure~\ref{f5} we show the 
dependence on these parameters of the location of the stagnation point 
$\mathcal{S}$ (cf.~Eq.~\ref{e2.23}), the maximum velocity attained by the 
ejected material $V_\mathrm{ej}$ (cf.~Eq.~\ref{e2.33}), and the contrast 
$\delta$ between the polar and equatorial densities at the injection sphere 
(cf.~Eq.~\ref{e2.38}).

From this figure we see that as $\dot{M}_\mathrm{in}/\dot{M}_\mathrm{M}$ 
increases the stagnation point sinks closer to the central accretor, while at 
the same time the velocity of the ejected material approaches the speed of 
light and the density contrast increases. From \eq{e2.52} and Figure~\ref{f14} 
it is also clear that as the injection rate increases, more and more material 
is expelled from the system as a bipolar outflow. Moreover, the restrictions on 
the model parameters, as established in \eqs{e2.32} and \eqref{e2.51}, are also 
apparent in Figure~\ref{f5}: as soon as these limits are exceeded, the model 
ceases to be well-defined within the whole domain $r<\mathcal{R}$.

The non-relativistic limit of an ultrarelativistic, stiff fluid corresponds to 
an incompressible fluid \citep{tejeda18}. This Newtonian counterpart of the 
present analytic model is discussed in \cite{ATH19}. Indeed, it is simple to 
verify that, in the limit in which $V_0\ll 1$ and $\mathcal{R}\gg M$, all of 
the equations derived in this section for the velocity field, the streamlines, 
and the different mass fluxes reduce to the expressions presented in 
\cite{ATH19}.

The analytic model that we have presented allows a transparent understanding of 
the physics involved in the choked accretion mechanism. However, generalizing 
this model to accommodate a more realistic equation of state becomes 
analytically intractable. We explore this generalization in the next section by 
means of numerical simulations. 

\setcounter{equation}{0}
\section{Numerical simulations}
\label{s3}

The main limitation of the analytic model for choked accretion discussed in the 
previous section is that it is based on the assumption of an ultrarelativistic 
gas with a stiff equation of state, an assumption with a rather restricted 
applicability in astrophysics. In this section we want to explore whether the 
phenomenon of choked accretion might also arise when considering more general 
equations of state, thus relaxing the associated assumption of a potential 
flow. This exploration will be based on full-hydrodynamic, numerical 
simulations performed with the open source code \textit{aztekas}.

The general relativistic hydrodynamic equations are solved numerically with 
\textit{aztekas} by recasting them in a conservative form using the Valencia 
formulation \citep{banylus}. The \textit{aztekas} code uses a grid based finite 
volume scheme, a High Resolution Shock Capturing method with an approximate 
Riemann solver for the flux calculation, and a monotonically centered second 
order reconstructor at cell interfaces. The code adopts a second order total 
variation diminishing Runge-Kutta method \citep{shu1988} for the time 
integration. See \cite{TA19} for further details about the discretization used 
in \textit{aztekas}. Code validation through comparisons to standard analytical 
solutions in the Newtonian and relativistic regimes can be found in 
\cite{aguayo18,TA19,ATH19}, while a number of standard shock tube tests 
successfully reproduced by the code are included in Appendix \ref{AB}.

The simulations presented in this section were performed for a perfect fluid 
evolving in a fixed, background metric corresponding to a Schwarzschild black 
hole of mass $M$. We adopt horizon-penetrating, Kerr-Schild coordinates and, 
imposing axisymmetry, we consider only 2D spatial domains with spherical 
coordinates $r$ and $\theta$. 

Furthermore, by assuming symmetry with respect to the equatorial plane located 
at $\theta=\pi/2$ (north-south symmetry), we restrict the numerical domain as 
$(r,\,\theta)\in[\mathcal{R}_\text{acc},\,\mathcal{R}]\times[0,\,\pi/2]$, where 
 $\mathcal{R}_\text{acc}$ is the radius of the inner boundary at which we adopt 
a free outflow condition (i.e.~free inflow onto the central black hole) and 
$\mathcal{R}$ is the radius of the injection sphere at which we impose a given 
profile for the physical parameters of the injected fluid. At both polar 
boundaries $\theta=0,\,\pi/2$ we adopt reflection conditions.

As initial conditions, we populate the whole numerical domain with the same 
values as those used at the outer boundary $\mathcal{R}$. For the code we adopt 
geometrized units, and take $M=1$ as unit of length and time. 

\subsection{Stiff fluid}
\label{s3.1}

 \begin{figure}
 \begin{center}
  \includegraphics[width=0.99\linewidth]{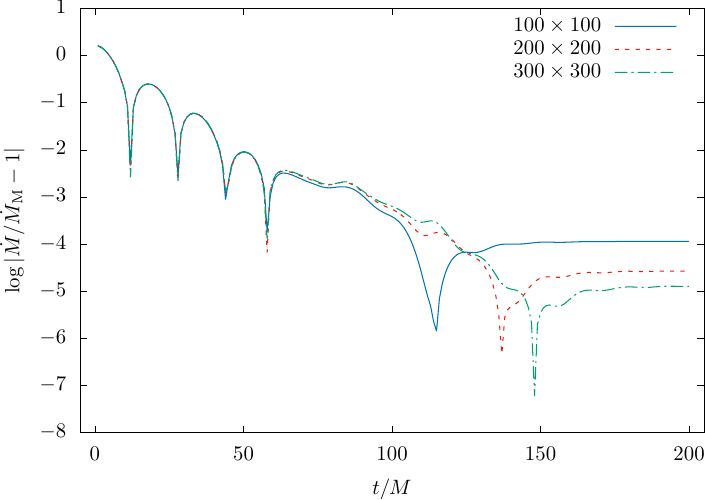} 
 \end{center}
\caption{Benchmark test of \textit{aztekas} with the analytic model of choked 
accretion described in Section~\ref{s2}. In this case we took $\mathcal{R}=10M$ 
as radius of the injection sphere and $V_0 = 0.16$ at the equator. The figure 
shows the time evolution of the relative error between the numerically 
calculated accretion rate $\dot{M}$ and the exact value of $\dot{M}_\mathrm{M} 
= 16\pi M^2\alpha_0\rho_0\Gamma_0$ for three resolutions (grid points) 
100$\times$100, 200$\times$200, and 300$\times$300. Note that the sharp falls 
observed in this figure correspond to changes in sign of the relative error 
being plotted. Moreover, the apparent periodicity observed at the beginning of 
the curve corresponds to an initial transient mode reflecting back and forth 
throughout the numerical domain at the speed of sound (in this case, the sound 
crossing time is $t \sim 10M$).}
\label{f6}
 \end{figure}

 \begin{figure*}
 \begin{center}
  \includegraphics[width=0.49\linewidth]{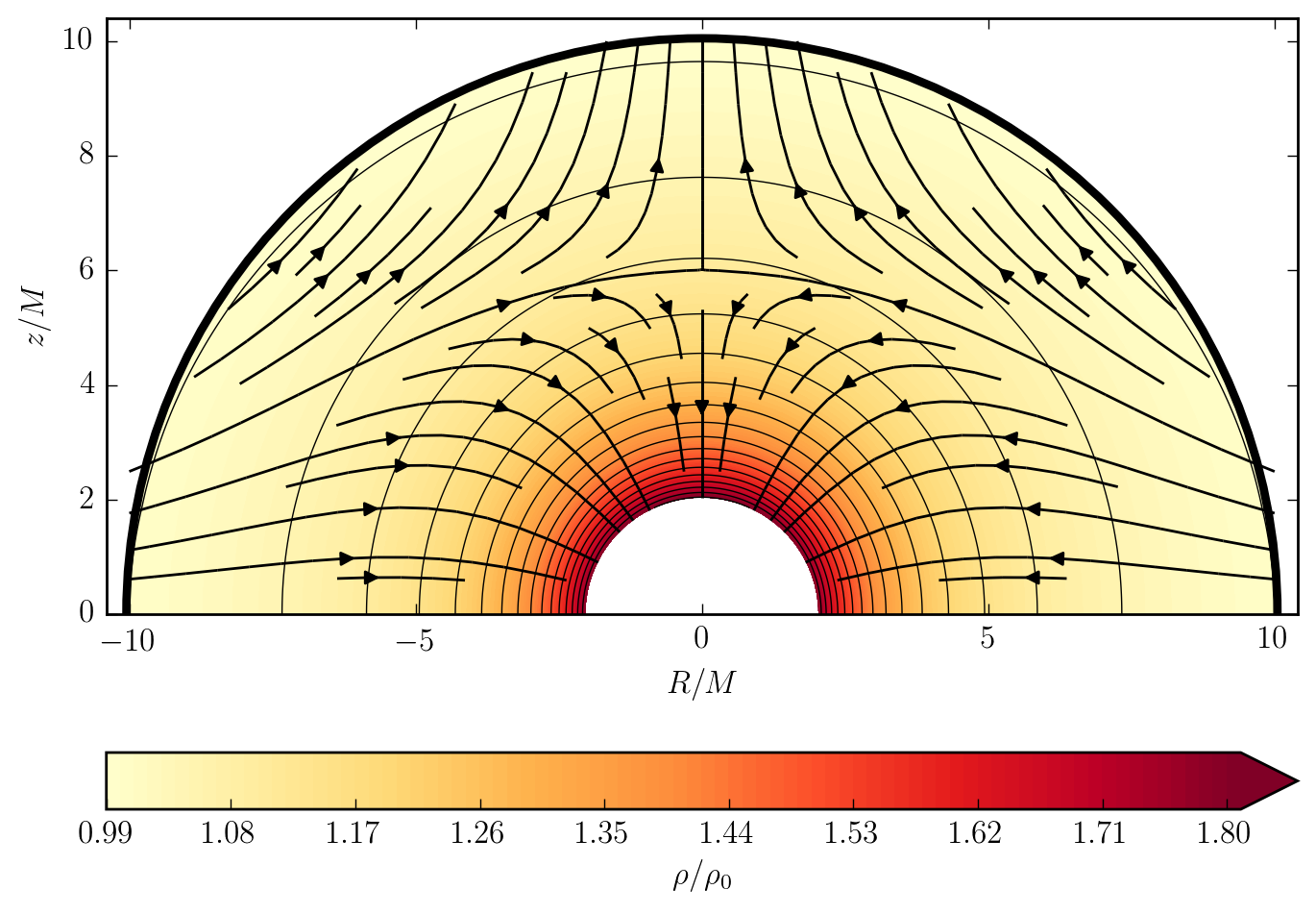} 
  \includegraphics[width=0.49\linewidth]{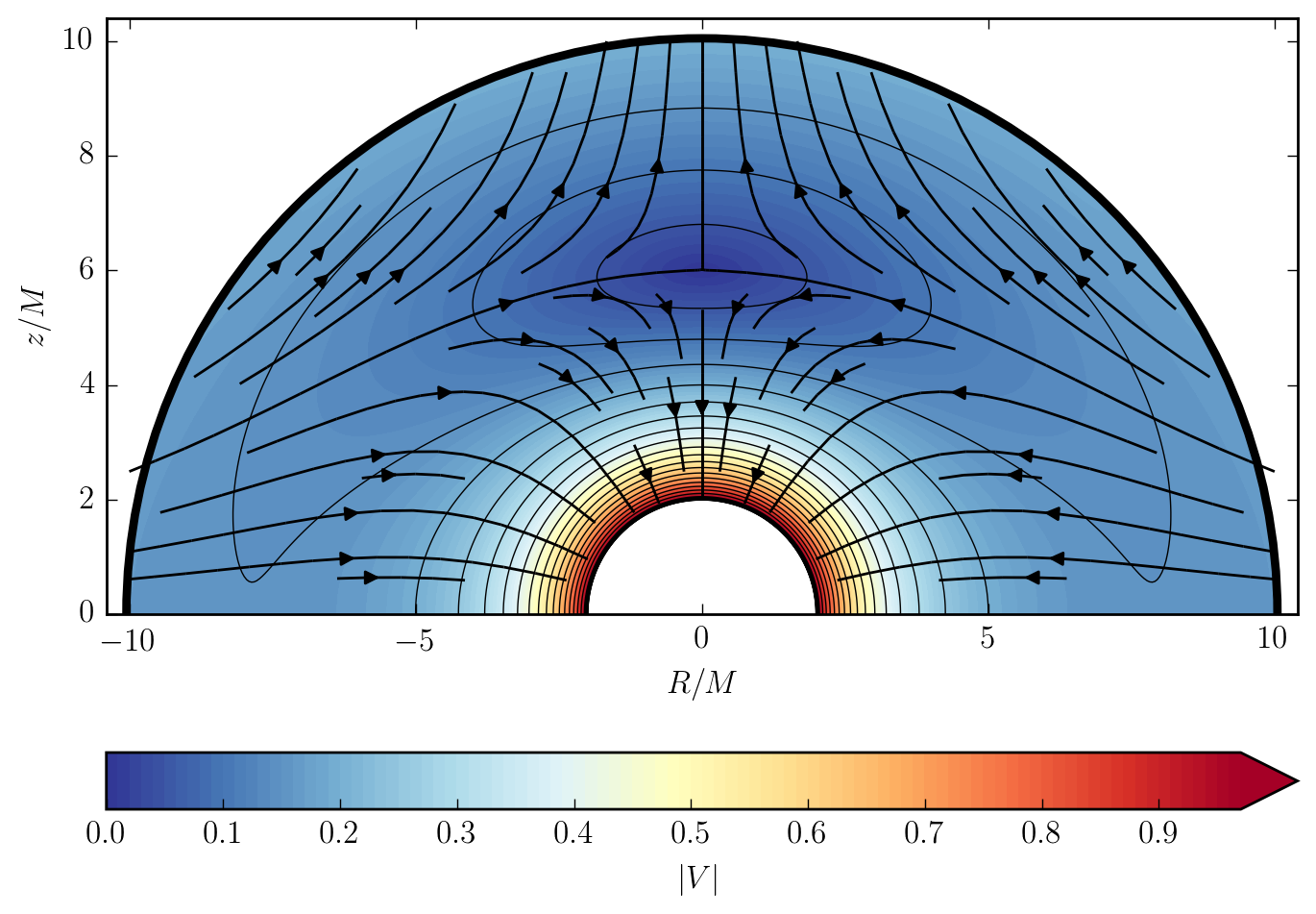} 
 \end{center}
\caption{Benchmark test of \textit{aztekas} with the analytic model of choked 
accretion described in Section~\ref{s2}. In this case we took $\mathcal{R}=10M$ 
as radius of the injection sphere and $V_0 = 0.16$ at the equator. The left 
panel shows isocontour levels of the density field with the scale indicated by 
the color bar. The right panel shows isocontour levels of the magnitude of the 
three-velocity. Fluid streamlines are indicated by thick, solid lines with an 
arrow. The simulation time is $t = 650M$.  An excellent agreement is found 
between this figure and its analytic counterpart in Figure~\ref{f4}. }
\label{f7}
 \end{figure*}

An analytic model can be useful as a benchmark solution for testing the ability 
of a numerical code to recover certain behaviour under appropriate conditions. 
Here we use the exact analytic solution presented in the previous section as a 
benchmark test for \textit{aztekas}. Specifically, we will consider the values 
of $\mathcal{R}=10M$  for the radius of the injection sphere and $V_0 = 0.16$ 
for the magnitude of the three-velocity at the equator of the injection sphere. 

For this test we take as radial boundaries $\mathcal{R}_\text{acc}=M$ and 
$\mathcal{R}=10M$ and use three different resolutions (grid points) 
100$\times$100, 200$\times$200, and 300$\times$300 for the radial and polar 
ranges. We adopt the approximation of an ultrarelativistic gas with a stiff 
equation of state as described in Section~\ref{s2}. At the injection sphere we 
impose the analytic value for the density as given in \eq{e2.36} and the 
velocity components corresponding to the transformation from Schwarzschild 
coordinates to Kerr-Schild coordinates.\footnote{The transformation between 
Schwarzschild ($t$, $r$, $\theta$, $\phi$) and  Kerr-Schild coordinates ($T$, 
$r$, $\theta$, $\phi$) is given by $$ \ud T = \ud t + \frac{2M}{r-2M}\,\ud r, 
$$ while the spatial components remain unchanged. The radial and polar 
components of the four-velocity in Kerr-Schild coordinates are then given by 
\eqs{e2.18} and \eqref{e2.19}, whereas the time component is now given by
$$ \frac{h}{e}\frac{\ud T}{\ud \tau}  = 
\left(1-\frac{2M}{r}\right)^{-1}\left(1+\frac{2M}{r} 
\frac{h}{e}\frac{\ud r}{\ud \tau}\right).$$ }

We let the simulations run until a steady-state condition is reached. This is 
monitored by calculating the mass accretion rate $\dot{M}$ across 
$\mathcal{R}_\text{acc}$. In Figure~\ref{f6} we show the time evolution of 
the numerically calculated $\dot{M}$ as compared to the analytically expected 
value of $\dot{M}_\mathrm{M} = 16\pi M^2\alpha_0\rho_0\Gamma_0$ for the three 
adopted resolutions. As can be seen from this figure, the value of $\dot{M}$ 
rapidly stabilizes to a constant value that agrees with $\dot{M}_\mathrm{M}$ to 
within $0.001\%$ for the largest resolution considered.

In Figure~\ref{f7} we show the density and velocity fields of the 
\textit{aztekas} simulation at $t=650M$. The stagnation point in the numerical 
simulation is located at $\mathcal{S} = 6.0175 M$, which is consistent with the 
analytically exact value of $\mathcal{S} = 6M$, taking into account the radial 
grid size of $\Delta r = 0.045 M$. 

We have also tried this benchmark test with different values of $\mathcal{R}$ 
and $V_0$, and consistently found that the numerical results recover the 
analytic solution in this limit case of a stiff equation of state, thus 
validating our numerical setup.

\subsection{Polytropic fluids}

In this section we relax the stiff fluid condition and consider perfect fluids 
described by a polytropic equation of state of the form 
\begin{equation}
 P = K\,\rho^\gamma,
 \label{e3.1}
\end{equation}
with $\rho$ the rest mass density, $\gamma$ the adiabatic index, and $K=\co$  
From \eq{e2.3}, the sound speed corresponding to this equation of state is 
given by
\begin{equation}
 a^2 = \frac{\gamma}{ h }\frac{P}{\rho},
 \label{e3.2}
\end{equation}
where $ h $ is the relativistic specific enthalpy. For a perfect fluid 
described by \eq{e3.1}, $ h $ is related to the other thermodynamical variables 
through
\begin{equation}
  h  = 1 + \frac{\gamma}{\gamma - 1}\frac{P}{\rho},
 \label{e3.3}
\end{equation}
or, by combining \eqs{e3.2} and \eqref{e3.3}, we can also write
\begin{equation}
  h  = \frac{1}{1-a^2/(\gamma-1)}.
 \label{e3.4}
\end{equation}

The expected requirement for the appearance of choked accretion, is that there 
should exist a small contrast between the density at the equator and that at 
the poles of the injection sphere. Here we impose this density contrast by 
adopting the following density profile as boundary condition at the injection 
radius 
$\mathcal{R}$
\begin{equation}
 \rho(\theta) = \rho_0 \left(1-\delta\cos^2\theta\right),
 \label{e3.5}
\end{equation}
where $\rho_0$ is the value of the density at the equator of the injection 
sphere, i.e~$\rho_0 = \rho(\pi/2)$, and $\delta$ is the same density contrast 
between the equator and the poles as defined in \eq{e2.38}.

The specific functional form of the boundary condition in \eq{e3.5} was chosen 
as a convenient first-order parametrization of a bipolar deviation from 
spherical symmetry. This profile is qualitatively similar to the one of the 
analytic model presented above (see bottom panel of Figure~\ref{f4}). We have 
explored with other similar boundary profiles and obtained consistent results.

For the simulations reported in this work we have taken \mbox{$\rho_0 = 
10^{-10}$}, although we have found that taking any other value of $\rho_0$ 
results in the same flow structure but with the density re-scaled by this new 
factor. In other words, the value of $\rho_0$ can be set arbitrarily, thus 
defining a unit scale for the density and related thermodynamical quantities, 
provided the fluid considered remains a negligible perturbation on the 
background metric. On the other hand, the resulting steady state solution 
depends strongly on the value of the sound speed $a_0$ imposed at the equator 
of the injection sphere. Note in particular that, from \eq{e3.4} and for a 
fixed adiabatic index $\gamma$, $a_0$ is limited as
\begin{equation}
 0 < a_0 < \sqrt{\gamma-1}.
 \label{e3.6}
\end{equation}

Once we have adopted a given $\gamma$,  set up values for $\rho_0$ and $a_0$ at 
the equator of the injection sphere and a density contrast $\delta$, we use the 
equation of state in \eq{e3.1} to find the corresponding pressure profile at 
the 
injection boundary as
\begin{equation}
 P(\theta) = \frac{1}{\gamma}\left[ \frac{a^2_0}{1-a^2_0/(\gamma - 
1)}\right]\frac{\rho(\theta)^\gamma}{\rho_0^{\gamma-1}}.
\label{e3.7}
 \end{equation}
 
Since we do not know the structure of the accretion flow beforehand (on which 
no a priori restrictions are imposed), we can not prescribe specific values for 
the velocity components at the injection radius.  For this reason we adopt 
free-boundary conditions for the radial and polar components of the velocity 
field and let the simulation evolve starting off from an initial state at rest 
(zero initial velocities), until an equilibrium state is reached throughout the 
numerical domain. Note that this means that we can not use the same 
parametrization that we had adopted for the analytic model of Section~\ref{s2} 
(i.e. $\mathcal{R}/M$, $\rho_0$, $P_0$, and $V_0$), but instead now we shall 
replace $V_0$ by the density contrast $\delta$.

\subsubsection{Dependence on the density contrast $\delta$}

\begin{figure*}
 \begin{center}
  \includegraphics[width=0.49\linewidth]{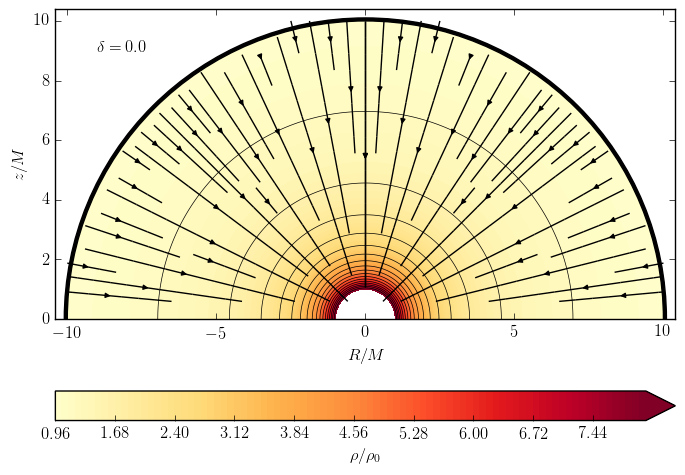} 
  \includegraphics[width=0.49\linewidth]{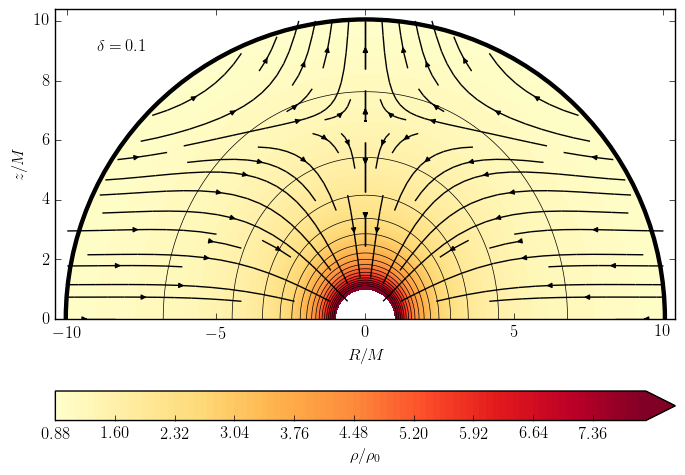} \\
  \includegraphics[width=0.49\linewidth]{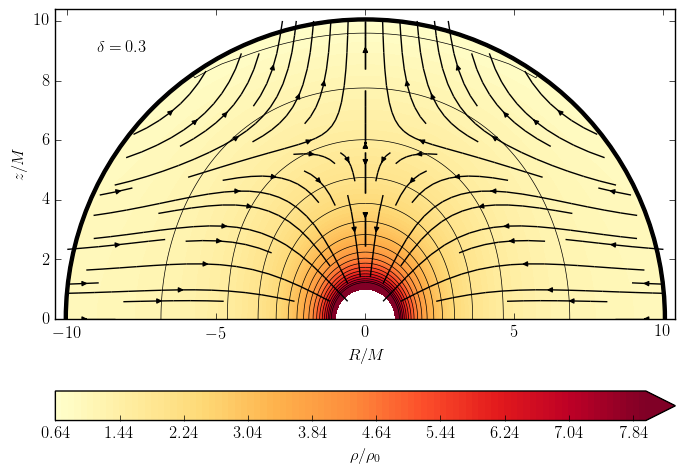} 
  \includegraphics[width=0.49\linewidth]{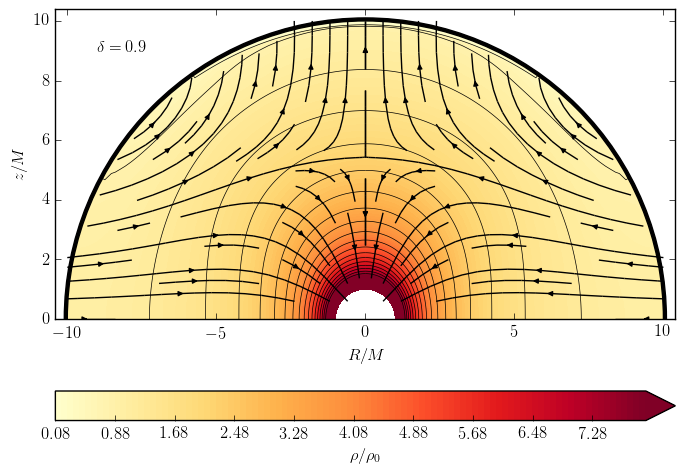} 
 \end{center}
\caption{Resulting steady state flow configuration for the numerical 
simulations for a polytropic fluid with $\gamma = 4/3$ accreting onto a 
Schwarzschild black hole. The value of the density contrast $\delta$ used in 
each case is indicated on the top-left corner of each panel, and increases 
gradually from $\delta = 0$ in the first panel (isotropic case where the Michel 
solution is recovered) to the highly anisotropic $\delta=0.9$ case in the 
fourth 
panel.}
\label{f8}
 \end{figure*}

Based on the analytic results of Section~\ref{s2}, we expect ejection rates and 
velocities to strongly correlate with the density contrast at the injection 
surface. Here we study the role of the density contrast as parametrized by 
$\delta$ in \eq{e3.5}. In Figure~\ref{f8} we show the steady-state results for 
four numerical simulations with $\gamma=4/3$, $a_0 = 0.5$, $\mathcal{R} = 10M$, 
and density contrasts $\delta=0,\,0.1,\,0.3$ and 0.9. The results of these four 
simulations, including also the $\delta = 0.5$ case, are reported in 
Table~\ref{t_delta}.  We can see that, as soon as the equatorial region becomes 
over dense with respect to the polar regions, i.e.~$\delta>0$, a strong 
qualitative change ensues with an inflow-outflow configuration appearing across 
the numerical domain. The resulting streamlines closely resemble the flow 
morphology of the analytic model presented in the previous section.\footnote{We 
note that since the posting of the initial version of this paper, a couple of 
relevant independent results have appeared: \cite{waters2020} present a 
numerical scheme using the ATHENA++ code simulating accretion onto a black 
hole, modeled using a pseudo-Newtonian potential, which yields very similar 
results to what we obtain, for the same angular accretion density profile which 
we present. \cite{zeraa2020} explore a similar scenario through an approximate 
semi-analytic approach, including the additional physical ingredients of 
rotation, viscosity and radiation pressure, to again obtain flow patterns 
highly resembling our results, provided an equatorial to polar accretion 
density profile is present.}

We can also see that as the density contrast increases, both $\dot{M}$ and 
$\dot{M}_\mathrm{in}$ increase. Note however, that $\dot{M}_\mathrm{in}$ 
increases faster than $\dot{M}$, with the net result that the ratio between 
ejection and injection also increases with increasing $\delta$. 

As a further positive test of our numerical scheme, when $\delta=0$ the 
simulation recovers the analytic solution of spherically symmetric accretion 
discussed by \cite{michel72}. The sixth column in Table~\ref{t_delta} reports 
the ratio $\dot{M}/\dot{M}_\mathrm{M}$, i.e.~the numerically found mass 
accretion rate in units of the mass accretion rate of Michel's solution 
$\dot{M}_\mathrm{M}$ (an analytic expression for $\dot{M}_\mathrm{M}$ is 
derived in  Appendix~\ref{sA}). Note that, as occurs also in the analytic model 
discussed in the previous section, this ratio does not remain strictly equal to 
one as $\delta$ increases, although the mass accretion rate remains of the 
order of $\dot{M}_\mathrm{M}$.

This first exploration confirms that the basic principle behind the choked 
accretion model presented in Section~\ref{s2} also works for fluids described 
by more general equations of state, where the resulting flow pattern is no 
longer assumed to be a potential flow.
 
\begin{deluxetable}{ccccccccc}
\tablecaption{Dependence on the density contrast $\delta$ \label{t_delta}}
\tablewidth{0pt}
\tablehead{ $\delta$ & $\dot{M}$ &  $\dot{M}_\mathrm{in}$ & 
$\dot{M}_\mathrm{ej}$ & 
$\frac{\dot{M}_\mathrm{ej}}{\dot{M}_\mathrm{in}}$ & 
$\frac{\dot{M}}{\dot{M}_\mathrm{M}}$ & $\mathcal{S}$ & $V_\mathrm{ej}$
}
\startdata
0.0 &  9.08 & 9.08  & 0.0  & 0.0  & 1.0  & ...   &  ... \\[2pt]
0.1 & 9.55 & 11.16 & 1.61 & 0.14 & 1.05 & 6.63 & 0.25 \\[2pt]
0.3 & 10.35 & 14.10 & 3.75 & 0.27 & 1.14 & 5.91 & 0.43 \\[2pt]
0.5 & 11.14 & 16.24 & 5.10 & 0.31 & 1.23 & 5.64 & 0.46 \\[2pt]
0.9 & 12.81 & 19.77 & 6.96 & 0.35 & 1.41 & 5.46 & 0.47 \\
\enddata
\tablecomments{The simulation parameters are fixed as $\mathcal{R} = 10M$, 
$\gamma=4/3$  and $a_0 = 0.5$. All the accretion rates are expressed in units 
of $\dot{M}_0 = M^2\rho_0$, and the stagnation point in units of $M$. The 
velocity $V_\mathrm{ej}$ is defined as the magnitude of the three-velocity at 
the poles of the injection sphere. According to \eq{ea.15}, the Michel mass 
accretion rate in this case is given by $\dot{M}_\mathrm{M} = 9.08 \dot{M}_0$.  
}
\end{deluxetable}

\subsubsection{Dependence on the adiabatic index $\gamma$}

\begin{figure}
 \begin{center}
  \includegraphics[width=0.98\linewidth]{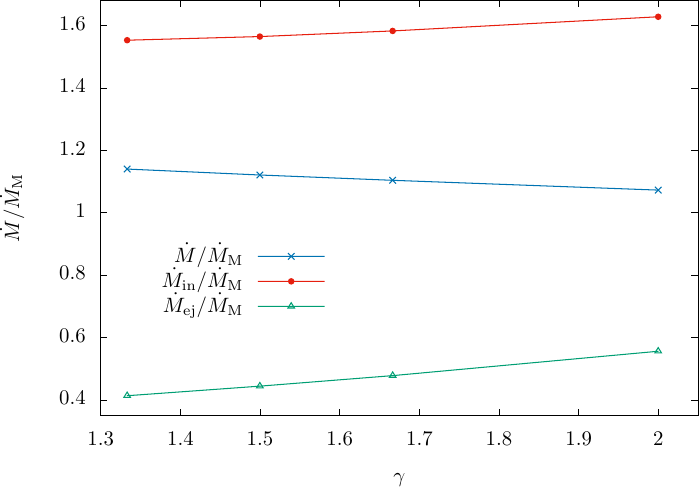}
 \end{center}
\caption{Dependence of the different mass flux rates (in units of the 
corresponding Michel value $\dot{M}_\mathrm{M}$) on the polytropic index 
$\gamma$.}
\label{f9}
 \end{figure}

Here we examine the behaviour of the steady-state, numerical solution as a 
function of the adiabatic index $\gamma$. We keep as fixed parameters 
$\mathcal{R}=10M$, $\delta=0.3$ and $a_0 = 0.5$ while considering four different 
values of $\gamma = 4/3,\,3/2,\,5/3$, and 2. In Table~\ref{t_gamma} we summarize 
the results of these simulations, while in Figure~\ref{f9} we plot the ratios 
$\dot{M}/\dot{M}_\mathrm{M}$, $\dot{M}_\mathrm{in}/\dot{M}_\mathrm{M}$ and 
$\dot{M}_\mathrm{ej}/\dot{M}_\mathrm{M}$ as functions of $\gamma$. The resulting 
density field and fluid streamlines for these four simulations are qualitatively 
similar to those shown in the bottom left panel of Figure~\ref{f8}. 

\begin{deluxetable}{ccccccccc}
\tablecaption{Dependence on  the adiabatic index $\gamma$ \label{t_gamma}}
\tablewidth{0pt}
\tablehead{ $\gamma$ & $\dot{M}$ &  $\dot{M}_\mathrm{in}$ & 
$\dot{M}_\mathrm{ej}$ & 
$\frac{\dot{M}_\mathrm{ej}}{\dot{M}_\mathrm{in}}$ & 
$\frac{\dot{M}}{\dot{M}_\mathrm{M}}$ & $\mathcal{S}$ & $V_\mathrm{ej}$
}
\startdata
4/3 & 10.35 & 14.10 & 3.75 & 0.27 & 1.14 & 5.91 & 0.43 \\[2pt]
3/2 & 9.75 & 13.61 & 3.86 & 0.28 & 1.12 & 5.82 & 0.42 \\[2pt]
5/3 & 9.15 & 13.12 & 3.96 & 0.30 & 1.10 & 5.73 & 0.41 \\[2pt]
2 & 8.01 & 12.16 & 4.15 & 0.34 & 1.07 & 5.55 & 0.40 \\
\enddata
\tablecomments{The simulation parameters are $\mathcal{R}=10M$, $a_0 = 0.5$ and 
$\delta=0.3$. From top to bottom the values of $\dot{M}_\mathrm{M}/\dot{M}_0$ 
are 9.08, 8.70, 8.28 and 7.47.  }
\end{deluxetable}

\begin{deluxetable}{ccccccccc}
\tablecaption{Dependence on the density contrast $\delta$  for $a_0 = 0.2$ 
\label{t_a1}}
\tablewidth{0pt}
\tablehead{ $\delta [\%]$ & $\dot{M}$ &  $\dot{M}_\mathrm{in}$ & 
$\dot{M}_\mathrm{ej}$ & 
$\frac{\dot{M}_\mathrm{ej}}{\dot{M}_\mathrm{in}}$ & 
$\frac{\dot{M}}{\dot{M}_\mathrm{M}}$ & $\mathcal{S}$ & $V_\mathrm{ej}$
}
\startdata
0.1 & 63.12 & 70.50 & 7.37 & 0.10 & 0.98 & 66.28 & 0.014 \\[2pt]
0.5 & 63.93 & 89.90 & 25.97 & 0.29 & 0.99 & 56.78 & 0.025 \\[2pt]
1.0 & 64.13 & 97.61 & 33.49 & 0.34 & 1.00 & 51.08 & 0.031 \\[2pt]
5.0 & 64.78 & 159.93 & 95.15 & 0.59 & 1.01 & 39.68 & 0.071 \\
\enddata
\tablecomments{The sound speed $a_0$ is given at the equator of the injection 
sphere. The simulation parameters are $\mathcal{R}=100M$ and $\gamma = 5/3$. In 
this case $\dot{M}_\mathrm{M} = 64.39 \dot{M}_0$. }
\end{deluxetable}

\begin{deluxetable}{ccccccccc}
\tablecaption{Dependence on the density contrast $\delta$  for $a_0 = 0.4$ 
\label{t_a2}}
\tablewidth{0pt}
\tablehead{ $\delta [\%]$ & $\dot{M}$ &  $\dot{M}_\mathrm{in}$ & 
$\dot{M}_\mathrm{ej}$ & 
$\frac{\dot{M}_\mathrm{ej}}{\dot{M}_\mathrm{in}}$ & 
$\frac{\dot{M}}{\dot{M}_\mathrm{M}}$ & $\mathcal{S}$ & $V_\mathrm{ej}$
}
\startdata
0.1 & 17.49 & 43.25 & 25.76 & 0.60 & 1.10 & 42.53 & 0.019 \\[2pt]
0.5 & 17.14 & 79.47 & 62.33 & 0.78 & 1.08 & 33.03 & 0.043 \\[2pt]
1.0 & 16.49 & 106.08 & 89.60 & 0.84 & 1.04 & 29.23 & 0.061 \\[2pt]
5.0 & 17.74 & 222.67 & 204.93 & 0.92 & 1.11 & 22.58 & 0.136 \\
\enddata
\tablecomments{The sound speed $a_0$ is given at the equator of the injection 
sphere. The simulation parameters are $\mathcal{R}=100M$ and $\gamma = 5/3$. In 
this case $\dot{M}_\mathrm{M} = 15.93 \dot{M}_0$. }
\end{deluxetable}

\begin{deluxetable}{ccccccccc}
\tablecaption{Dependence on the density contrast $\delta$  for $a_0 = 0.6$ 
\label{t_a3}}
\tablewidth{0pt}
\tablehead{ $\delta [\%]$ & $\dot{M}$ &  $\dot{M}_\mathrm{in}$ & 
$\dot{M}_\mathrm{ej}$ & 
$\frac{\dot{M}_\mathrm{ej}}{\dot{M}_\mathrm{in}}$ & 
$\frac{\dot{M}}{\dot{M}_\mathrm{M}}$ & $\mathcal{S}$ & $V_\mathrm{ej}$
}
\startdata
0.1 & 10.82 & 52.63  & 41.81  & 0.79 & 1.32 & 33.03 & 0.029 \\[2pt]
0.5 & 11.35 & 108.20 & 96.85  & 0.90 & 1.39 & 25.43 & 0.064 \\[2pt]
1.0 & 10.37 & 147.74 & 137.37 & 0.93 & 1.27 & 22.58 & 0.091 \\[2pt]
5.0 & 10.76 & 318.85 & 308.09 & 0.97 & 1.32 & 17.83 & 0.202 \\
\enddata
\tablecomments{The sound speed $a_0$ is given at the equator of the injection 
sphere. The simulation parameters are $\mathcal{R}=100M$ and $\gamma = 5/3$. In 
this case $\dot{M}_\mathrm{M} = 8.17 \dot{M}_0$. }
\end{deluxetable}

\begin{deluxetable}{ccccccccc}
\tablecaption{Dependence on the density contrast $\delta$  for $a_0 = 0.8$ 
\label{t_a4}}
\tablewidth{0pt}
\tablehead{ $\delta [\%]$ & $\dot{M}$ &  $\dot{M}_\mathrm{in}$ & 
$\dot{M}_\mathrm{ej}$ & 
$\frac{\dot{M}_\mathrm{ej}}{\dot{M}_\mathrm{in}}$ & 
$\frac{\dot{M}}{\dot{M}_\mathrm{M}}$ & $\mathcal{S}$ & $V_\mathrm{ej}$
}
\startdata
0.1 & 7.77 & 64.80  & 57.04  & 0.88 & 1.41 & 28.28 & 0.038 \\[2pt]
0.5 & 9.20 & 139.98 & 130.78 & 0.93 & 1.67 & 21.63 & 0.085 \\[2pt]
1.0 & 9.20 & 194.61 & 185.40 & 0.95 & 1.67 & 19.73 & 0.120 \\[2pt]
5.0 & 8.14 & 422.31 & 414.17 & 0.98 & 1.48 & 14.98 & 0.268 \\
\enddata
\tablecomments{The sound speed $a_0$ is given at the equator of the injection 
sphere. The simulation parameters are $\mathcal{R}=100M$ and $\gamma = 5/3$. In 
this case $\dot{M}_\mathrm{M} = 5.50 \dot{M}_0$. }
\end{deluxetable}

From these results we see a weak dependence on the adiabatic index $\gamma$. As 
we consider increasing values of $\gamma$, the values of $\dot{M}$,  
$\dot{M}_\mathrm{in}$, $\mathcal{S}$ and $V_\mathrm{ej}$ slightly decrease 
while both $\dot{M}_\mathrm{ej}$ and the ratio $\dot{M}_\mathrm{ej}/ 
\dot{M}_\mathrm{in}$ increase.

\subsubsection{Dependence on the sound speed $a_0$}

Now we turn our attention to the role played by the sound speed as defined at 
the equator of the injection sphere, $a_0$. We will also consider a larger 
injection radius than in the previous sections in order to probe a different 
regime with smaller density contrasts and larger mass injection rates. 
Specifically, we take $\mathcal{R} = 100M$, $\gamma = 5/3$, four density 
contrasts: $\delta =$ 0.1, 0.5, 1.0, 5.0\,\%, and four different values for the 
sound speed: $a_0=0.2$, 0.4, 0.6, and 0.8. Note that, from \eq{e3.6} for 
$\gamma = 5/3$, the maximum possible value for this parameter is $a_0=0.816$.

\begin{figure}
 \begin{center}
  \includegraphics[width=0.98\linewidth]{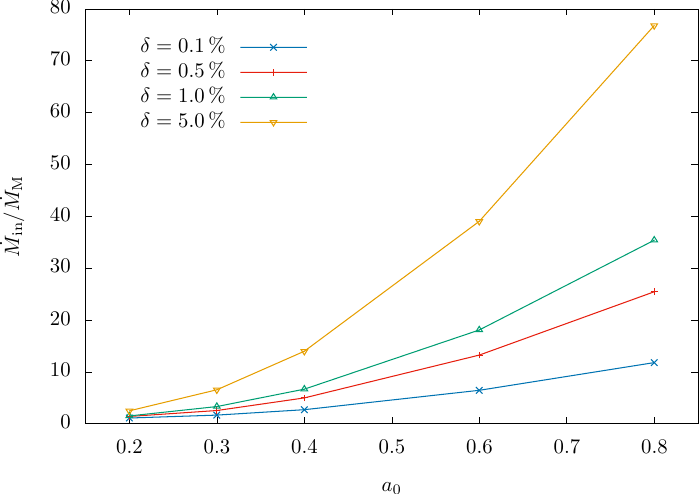}
 \end{center}
\caption{Dependence of the ratio $\dot{M}_\mathrm{in}/\dot{M}_\mathrm{M}$ on 
the sound speed $a_0$ as given at the equator of the injection sphere. Clearly 
this ratio is a monotonically increasing function of $a_0$ with a steeper 
growth with increasing $\delta$.}
\label{f10}
 \end{figure}

\begin{figure}
 \begin{center}
  \includegraphics[width=0.98\linewidth]{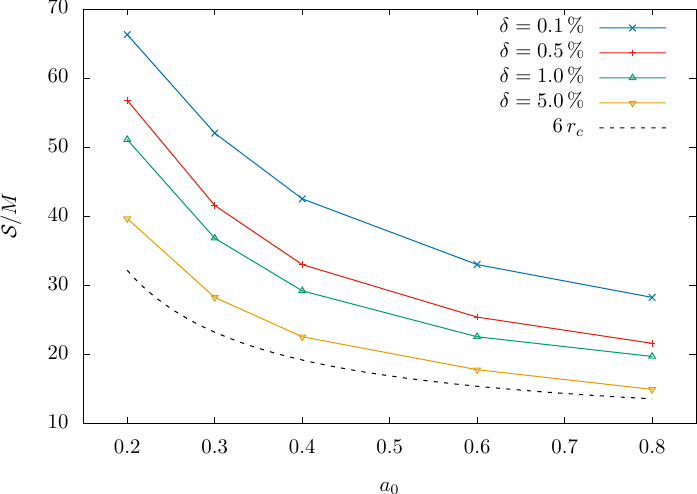}
 \end{center}
\caption{Dependence of the location of the stagnation point $\mathcal{S}$ on 
$a_0$. Here we see that $\mathcal{S}$ is inversely proportional to both $a_0$ 
and $\delta$. Note that $\mathcal{S}$ shows a dependence on $a_0$ that 
resembles the one followed by the critical radius $r_c$ on this same parameter.}
\label{f11}
 \end{figure}

\begin{figure}
 \begin{center}
  \includegraphics[width=0.98\linewidth]{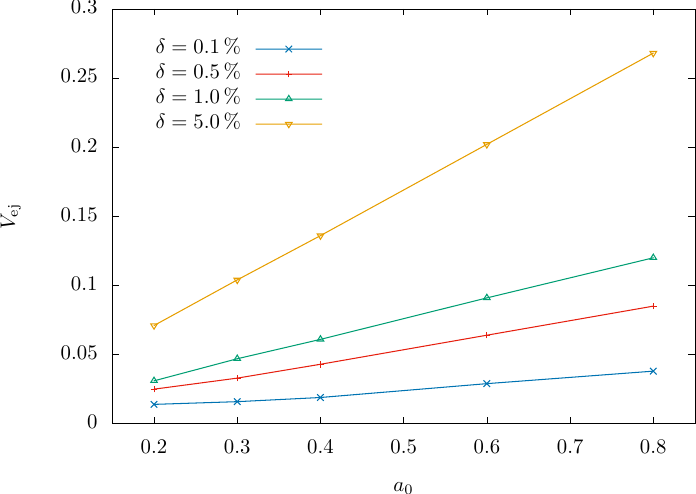}
 \end{center}
\caption{Dependence of the maximum velocity of the ejected material 
$V_\mathrm{ej}$ on $a_0$. Here we see that $V_\mathrm{ej}$ grows more or less 
linearly with $a_0$.}
\label{f12}
 \end{figure}

In Tables~\ref{t_a1}--\ref{t_a4}, we present a summary of the results obtained 
in this case. In Figure~\ref{f10} we show the dependence of the ratio 
$\dot{M}_\mathrm{in} /\dot{M}_\mathrm{M}$ on $a_0$ for the four values of the 
density contrast $\delta$. Figure~\ref{f11} shows the dependence of the 
location of the stagnation point $\mathcal{S}$ on $a_0$, while Figure~\ref{f12} 
shows the dependence of the maximum velocity of the ejected material 
$V_\mathrm{ej}$ on $a_0$. 

It is interesting to notice from Figure~\ref{f11} that, at least for the 
parameter space explored for this figure, $\mathcal{S}$ follows a dependence on 
$a_0$ similar to the one followed by the critical radius $r_c$ as defined in  
Appendix~\ref{sA} for the accretion flow in the spherically symmetric case.

\begin{figure*}
 \begin{center}
  \includegraphics[width=0.49\linewidth]{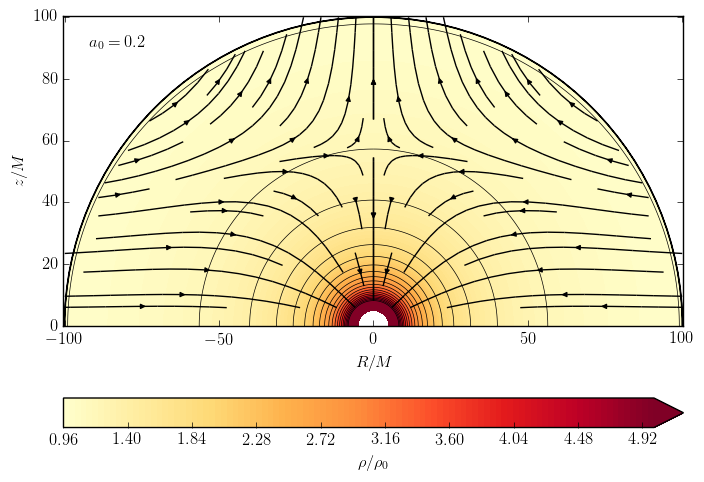}
  \includegraphics[width=0.49\linewidth]{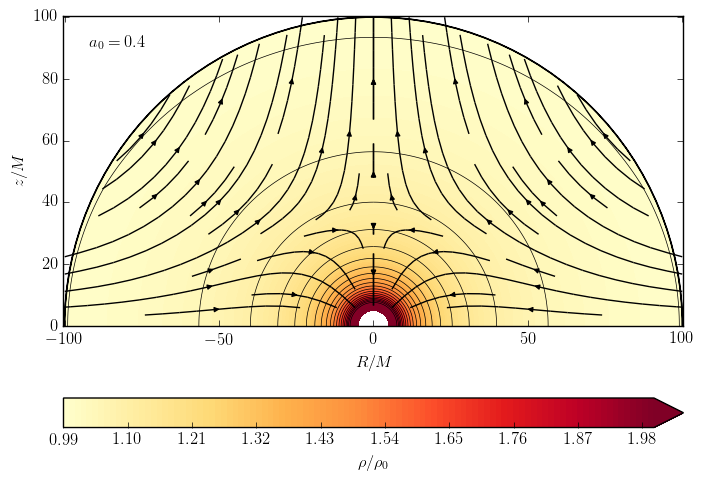}\\  
  \includegraphics[width=0.49\linewidth]{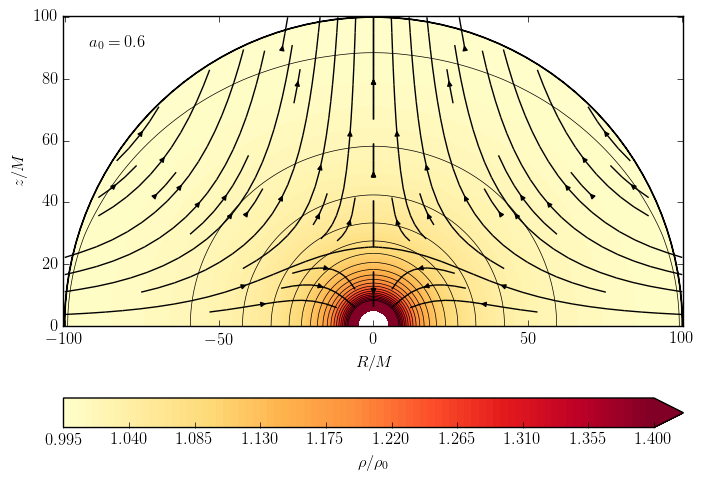}
  \includegraphics[width=0.49\linewidth]{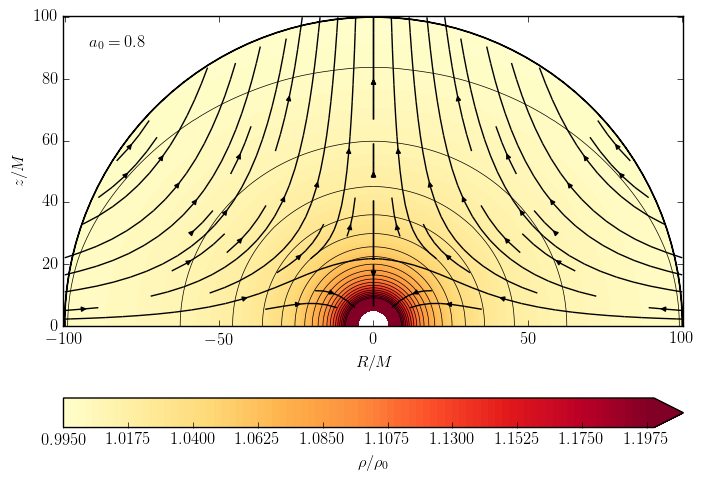}  
 \end{center}
\caption{Resulting steady state configurations for a polytropic fluid with 
$\gamma = 5/3$ and $\delta = 0.5\,\%$. The value of the sound speed $a_0$ used 
in each case is indicated on the top-left corner of each panel.}
\label{f13}
 \end{figure*}

In Figure~\ref{f13} we show the resulting steady flow configurations for the 
four values of $a_0$ in Tables~\ref{t_a1}--\ref{t_a4} and $\delta = 0.5\,\%$. 
The corresponding configurations for the other values of $\delta$ are 
qualitatively similar to the ones presented in this figure.

From these results we see that the final steady state configuration depends 
strongly on the value of the sound speed $a_0$. In general we see that as $a_0$ 
increases, the stagnation point $\mathcal{S}$ sinks deeper into the accretion 
flow, as more material is expelled from the system along the bipolar outflow at 
increasingly larger speeds $V_\mathrm{ej}$.  Moreover, we also see that as the 
influx asymmetry increases, even for a small $5\%$ density contrast, the 
ejection velocities become larger, reaching values of $V_\mathrm{ej} > 0.25$ 
for the sound speed values probed.

\setcounter{equation}{0}
\section{Discussion}
\label{s4}

\subsection{Comparison between the numerical simulations and the analytic model}

Even though the physics of the analytic model of Section~\ref{s2} differs from 
the one included in the simulations of Section~\ref{s3}, it is illustrative to 
compare the results of these two sections. For this comparison we will consider 
only the case of the $\gamma = 5/3$ polytrope and injection radius 
$\mathcal{R}=100M$ presented in Tables~\ref{t_a1}--\ref{t_a4}, as this large 
injection radius allowed us to explore a broader range of mass injection rates.

\begin{figure}
 \begin{center}
  \includegraphics[width=0.98\linewidth]{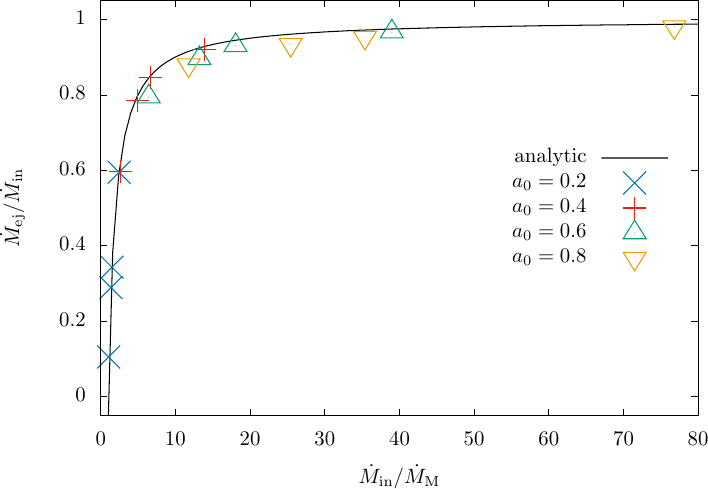}
 \end{center}
\caption{Ratio of ejected to injected mass rates 
$\dot{M}_\mathrm{ej}/\dot{M}_\mathrm{in}$ as a function of the injection mass 
rate in units of the Michel value $\dot{M}_\mathrm{M}$. The injection radius is 
$\mathcal{R}=100M$. The different symbols correspond to the numerical results 
reported in Tables~\ref{t_a1}--\ref{t_a4} for $\gamma = 5/3$ and sound speeds 
as labeled. The solid line corresponds to the analytic model of an 
ultrarelativistic $\gamma = 2$ stiff fluid presented in Section~\ref{s2} 
(cf.~Eq.~\ref{e2.52}).}
\label{f14}
 \end{figure}

In Figure~\ref{f14} we show the ratio of ejected over injected mass rates 
$\dot{M}_\mathrm{ej}/\dot{M}_\mathrm{in}$ as a function of the mass injection 
rate. From this figure we find a very good agreement between the numerical data 
and the analytic model. This agreement is remarkable if we take into account 
that the latter is based on the assumption of an ultrarelativistic stiff fluid, 
$\gamma = 2$, while the former involves a more realistic $\gamma = 5/3$ 
polytrope. We have also found this same agreement for different polytropic 
indices in the non-relativistic regime, as can be seen in Figure~7 of 
\cite{ATH19}.

\begin{figure}
 \begin{center}
  \includegraphics[width=0.98\linewidth]{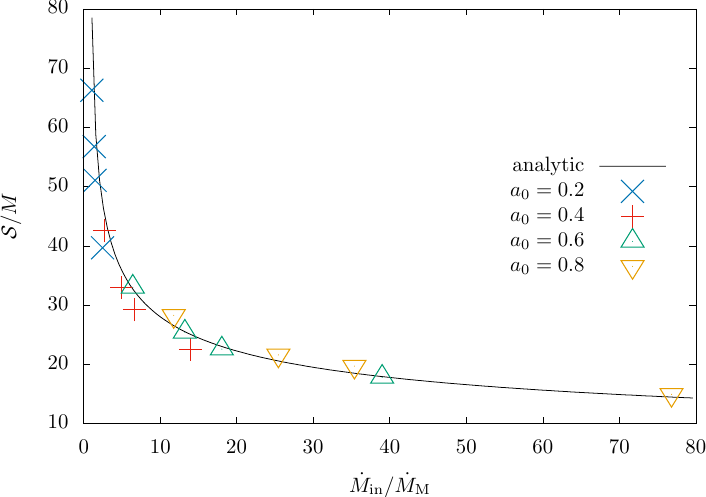}
 \end{center}
\caption{Location of the stagnation point $\mathcal{S}$ as a function of the 
injection mass rate. The injection radius is $\mathcal{R}=100M$. The different 
symbols correspond to the numerical results reported in 
Tables~\ref{t_a1}--\ref{t_a4} for $\gamma = 5/3$ and sound speeds as labeled. 
The solid line corresponds to the analytic model of an ultrarelativistic 
$\gamma = 2$ stiff fluid presented in Section~\ref{s2}.}
\label{f15}
 \end{figure}

In Figure~\ref{f15} we show the location of the stagnation point $\mathcal{S}$ 
as a function of the injection mass rate 
$\dot{M}_\mathrm{in}/\dot{M}_\mathrm{M}$. Note that as $\dot{M}_\mathrm{in}$ 
increases, $\mathcal{S}$ descends towards the central accretor, just as it 
occurred for the analytic model (cf. Figure~\ref{f5}). We find again a good 
agreement between the numerical data and the analytic model. 

\begin{figure}
 \begin{center}
  \includegraphics[width=0.98\linewidth]{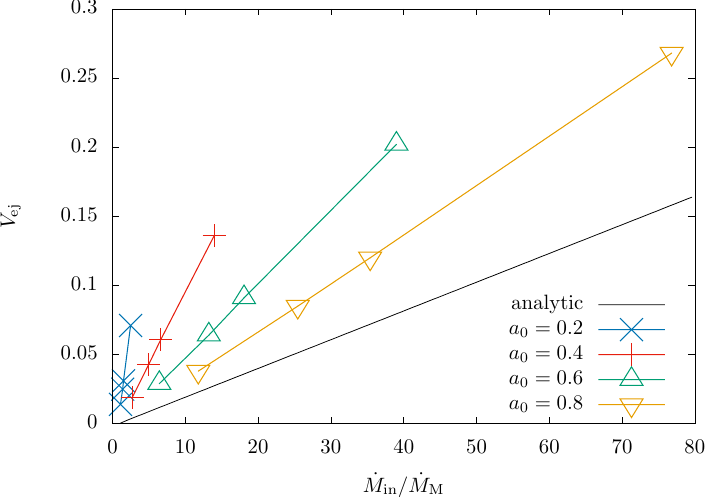}
 \end{center}
\caption{Maximum velocity attained by the ejected material $V_\mathrm{ej}$ as a 
function of the injection mass rate. The injection radius is 
$\mathcal{R}=100M$. The different symbols correspond to the numerical results 
reported in Tables~\ref{t_a1}--\ref{t_a4} for $\gamma = 5/3$ and sound speeds 
as labeled. The solid line corresponds to the analytic model of an 
ultrarelativistic $\gamma = 2$ stiff fluid presented in Section~\ref{s2}.
}
\label{f16}
 \end{figure}

In Figure~\ref{f16} we show the maximum velocity attained by the ejected 
material $V_\mathrm{ej}$ as a function of the injection mass rate 
$\dot{M}_\mathrm{in}/\dot{M}_\mathrm{M}$. We compare the numerical results 
against the analytic value for $V_\mathrm{ej}$ given in \eq{e2.31}. In contrast 
to what happens for the two parameters discussed above, here we find a large 
difference among the numerical results for each value of the sound speed $a_0$, 
as well as between these results and the analytic model. Note, however, that 
for each value of $a_0$ the numerically obtained values of $V_\mathrm{ej}$ 
follow a linear dependence on $\dot{M}_\mathrm{in}/\dot{M}_\mathrm{M}$ with a 
slope inversely proportional to $a_0$. Also, as $a_0$ increases the numerical 
data approaches the analytic model, for which $a = 1$ everywhere in the fluid.

\subsection{Applicability in astrophysics}

We discuss now the viability of the choked accretion phenomenon presented here 
for operating as the inner engine behind a given jet-launching astrophysical 
system. Given that the characteristic length scale of this mechanism is given 
by $\mathcal{S}$, we can expect the physical size of the inner accretion disk 
(that we have associated with $\mathcal{R}$) to be larger than $\mathcal{S}$. 
In general, for an accretion disk around a black hole, we will have 
$\mathcal{R}=1-10M$ (the actual value will be a function of both the disk model 
and the black hole spin). On the other hand, from all of the simulations 
presented in this work, as well as those in \cite{ATH19} for the 
non-relativistic case, we see that a robust lower limit for $\mathcal{S}$ is 
given by the corresponding Bondi radius $r_\mathrm{B}=M/a_\infty^2$. Moreover, 
provided that $\mathcal{R}>r_\mathrm{B}$, we have $a_0 \simeq a_\infty$ and 
then we can write 
\begin{equation}
\mathcal{S}>M/a_0^2 .
\label{e4.1}
\end{equation}

At this point it is useful to recall that, assuming an ideal gas, we can relate 
the  sound speed $a$ and the fluid temperature $T$ by
\begin{equation}
 \frac{T}{T_i}  = 
\left(\frac{\gamma-1}{\gamma}\right)\left(\frac{a^2}{\gamma-1-a^2}\right),
 \label{e4.2}
\end{equation}
where $T_i = m_i/k_B$ is the temperature corresponding to the rest mass energy 
of the average gas particle of mass $m_i$. For a gas composed of ionized 
hydrogen we have $T_H=1.08\times10^{13}\,\mathrm{K}$, while for an 
electron-positron plasma $T_e=5.93\times10^{9}\,\mathrm{K}$.

Then, from \eqs{e4.1} and \eqref{e4.2} we have
\begin{equation}
\frac{\mathcal{S}}{M}> \frac{1}{\gamma}\frac{T_i}{T}+\frac{1}{\gamma-1}.
\label{e4.3}
\end{equation}

A regular plasma dominated by radiation pressure and consisting of protons and 
electrons can be modeled, in a first approximation, as a $\gamma = 4/3$ 
polytrope with an average particle mass $m_i \simeq m_H$ and, thus, $T_i \simeq 
T_H$. As discussed in \cite{ATH19},  taking $T=10^7\,\mathrm{K}$ as the 
temperature at the inner edge of the disk in an X-ray binary \citep{xray}, from 
\eq{e4.3} we have $\mathcal{S} > 10^5 M$. In the case of an AGN, instead of 
taking the gas in the inner disk (at a temperature of around 
$T=10^5\,\mathrm{K}$) we can consider the ionized plasma in the hot corona 
above the disk with a temperature of up to $T=10^9\,\mathrm{K}$ ~\citep{agn}. 
Nevertheless, even for this large temperature, from \eq{e4.3} we obtain 
$\mathcal{S} > 10^3 M$. Even in the case of the accretion disk associated to a 
long GRB, where the gas temperature can reach up to 
$10^{11}\,\mathrm{K}$~\citep{woosley}, from \eq{e4.3} we have $\mathcal{S} > 80 
M$, which is still a factor of ten larger than the expected size of the inner 
engine.

The above analysis implies that, for the choked accretion mechanism to work for 
a regular plasma, the temperature of the infalling gas is required to be 
substantially higher than that of the inferred values at the inner edge of the 
disk (or disk corona). These higher temperatures could result from highly 
localized heating processes such as magnetic reconnection, shock heating, or 
viscous friction at the point of transition between the disk and the 
radial-infall domains.

On the other hand, the extreme conditions at the innermost part of these 
systems give rise to a different kind of plasma composed of relativistic 
electron-positron pairs \citep{Wardle98,Beloborodov99,Siegert16}. Considering 
that at least a fraction of this plasma has a thermal component, the pair 
production mechanism implies temperatures in excess of  $10^{11}\,\mathrm{K}$. 
If we consider this gas as the accreted material, then for this temperature and 
substituting $T_i = T_e$ in \eq{e4.3}, we get $\mathcal{S} >  M$. We obtain 
then that, under these circumstances, the choked accretion mechanism might 
become relevant for the ejection of this pair plasma. 

Contrary to the analytic model presented in Section~\ref{s2}, we do not have 
direct control on the mass injection rate crossing the outer boundary of our 
numerical simulations, as it is indirectly determined by the values of 
$\mathcal{R}$, $a_0$ and $\delta$. Nevertheless, it is clear that in an 
astrophysical scenario this mass rate will be imposed by external, possibly 
time varying conditions. For example, in the context of low mass X-ray 
binaries, stellar oscillations or orbital variations can modulate the total 
mass 
transfer across the Roche lobe from the regular star to the compact companion 
\citep{tauris06}. More dramatic time varying conditions will be found for jets 
launched during a common-envelope phase as studied by \cite{lopezcamara19}, or 
for long GRBs as studied by e.g.~\cite{lopezcamara10,TMP11}.

The strong dependence that we have found between the ratio of ejected to 
injected material and the incoming mass accretion rate, leads us to suggest 
that the choked accretion mechanism could offer a compelling, simple connection 
between the external mass flux feeding an accretion disk and the jet activity. 
Whenever the mass injection rate surpasses the threshold value 
$\dot{M}_\mathrm{M}$, the excess flux is prone to being ejected from the system 
as a bipolar outflow. This could be of relevance for studying the time 
variability of the jet emission. 

Once a tight connection appears between accretion rates and geometry on the one 
hand, and ejection rates and velocities on the other, we have the potential to 
correlate the time-variability in the mass flux across the accretion disk to 
the resulting ejection rates and velocities. This, in turn, naturally leads to 
the appearance of internal shocks in the ensuing jets, such as those typically 
assumed to be associated to the GRB phenomenology.

As already implied by the above discussion, a proper exploration of the role 
played by choked accretion in launching relativistic jets, demands accounting 
for additional physics, such as the effect of rotation, magnetic fields, and 
radiative transport. Indeed, these factors are considered as crucial for the 
acceleration and collimation of the resulting jets 
\citep{semenov04,McKinney06}. Moreover, as discussed in \cite{ATH19}, some of 
these ingredients might actually improve the applicability of choked accretion 
by increasing both the effective temperature and the polar density contrast, 
thus bringing $\mathcal{S}$ closer to the central accretor. It should also be 
interesting to study the possible interplay of choked accretion with the 
well-established \cite{BZ1977} mechanism. We intend to address these points in 
future work.

\setcounter{equation}{0}
\section{Summary}
\label{s5}

We have presented the choked accretion phenomenon as a purely hydrodynamical 
outflow-generating mechanism. Choked accretion operates under two basic 
premises: a sufficiently large mass flux accreting onto a central object, and 
an anisotropic density field in which an equatorial belt has a higher density 
than the polar regions. These two ingredients are plausibly met in several 
jet-launching astrophysical scenarios involving accretion discs around massive 
objects. We suggest that choked accretion constitutes a relevant ingredient for 
studying some of these systems. 

Moreover, we have shown that breaking spherical symmetry by imposing a polar 
density gradient in the accretion flow onto a central object, qualitatively 
changes the resulting steady state configurations from the purely radial 
accretion models \citep{bondi52,michel72}, to the infall-outflow morphology 
that characterizes the choked accretion model. Thus, choked accretion provides 
a natural transition between spherical accretion and systems characterized by 
bipolar outflows. 

We have studied this phenomenon by introducing first a general relativistic 
analytic model of choked accretion onto a Schwarzschild black hole. This model 
is based on the approximations of steady state, axisymmetry, and irrotational 
flow and assumes an ultrarelativistic stiff fluid.  We then relaxed this last 
assumption, together with the associated potential flow condition, and studied 
more general fluids by means of full-hydrodynamic simulations performed with 
the numerical code {\it aztekas}. 

Both for an ultrarelativistic stiff fluid as for a regular polytrope, the 
limiting value for the total mass accretion rate corresponds quite closely to 
the one found in the spherically-symmetric case \citep{michel72}.  We have thus 
found that, within the assumptions underlying this work, hydrodynamical 
accretion flows onto massive objects choke at this threshold value and any 
extra infalling material is deflected into a bipolar outflow.

The analytic solution presented here allowed us to study in detail the basic 
physical principle behind the choked accretion phenomenon. Moreover, we have 
also demonstrated the usefulness of this exact analytic solution as a benchmark 
test for validating numerical hydrodynamic codes in general. The 
non-relativistic limit of this analytic solution is presented in \cite{ATH19}.

Considering together: i) the perturbative Newtonian solutions for isothermal 
fluids of ~\cite{hernandez14}; ii) the exact Newtonian solution for 
incompressible fluids and the numerical Newtonian experiments for polytropic 
equations of state in \cite{ATH19}; and iii) the present exact analytic 
relativistic model for a stiff fluid and the numerical experiments presented 
for polytropic fluids; we can conclude that the inflow-outflow steady state 
configurations presented here are an extremely general and robust consequence 
of breaking spherical symmetry with a polar density gradient in an accretion 
flow onto a central object. Similarly, we see that the choked accretion 
character of these configurations extends across the Newtonian and relativistic 
regimes.

\section*{Acknowledgements}

We thank Olivier Sarbach and John Miller for insightful discussions and 
critical comments on the manuscript. We also thank Fabio de Colle and Diego 
L\'opez-C\'amara for useful comments and suggestions. This work was supported 
by DGAPA-UNAM (IN112616 and IN112019) and CONACyT (CB-2014-01 No.~240512; 
No.~290941; No.~291113) grants. AAO and ET acknowledge economic support from 
CONACyT (788898, 673583). XH acknowledges support from DGAPA-UNAM PAPIIT 
IN104517 and CONACyT.

\appendix

\setcounter{section}{0}
\setcounter{equation}{0}
\setcounter{figure}{0}

\renewcommand{\theequation}{A.\arabic{equation}}

\section{ Relativistic, spherically symmetric accretion flow }
\label{sA}
In this Appendix we give a brief overview of  \citet{michel72}'s analytic model 
of a spherically symmetric accretion flow onto a Schwarzschild black hole. In 
particular we derive an analytic expression for the resulting mass accretion 
rate in a form that is useful for the present work \citep[see][for an 
alternative derivation]{beskin95}. 

Here we will consider only the case of a perfect fluid described by a 
polytropic 
equation of state as in \eq{e3.1}. See \cite{chaverra15} for a recent 
extension of \citet{michel72}'s model to a general class of static, spherically 
symmetric background metrics, as well as for more  general equations of state.

Under the assumptions of stationary state and spherical symmetry, the equations 
governing the accretion flow are the continuity equation and the radial 
component of the relativistic Euler equation (Eqs.~\ref{e2.4} and \ref{e2.5}, 
respectively), i.e.
\begin{gather}
\frac{\ud }{\ud r}\left(r^2\rho\,U^r\right) = 0,
\label{ea.1}\\
\frac{\ud }{\ud r}\left(r^2T^r_t\right)= 0,
\label{ea.2}
\end{gather}
where $v = U^r=\ud r/\ud \tau$ and $T^r_t = \rho\,h\,U_t\,U^r$. Direct 
integration of these two equations gives
\begin{gather}
4\pi\,r^2\rho\,v = \dot{M}_\mathrm{M} = 4\pi\,\lambda =\co,
\label{ea.3}\\
\rho\,h\,U_t\,v\,r^2= \mu = \co
\label{ea.4}
\end{gather}
We can simplify \eq{ea.4} by dividing it by \eq{ea.3} and taking its square; 
the result is
\begin{equation}
h^2\left(1-\frac{2M}{r}+v^2\right)=\left(\frac{\mu}{\lambda}
\right)^2=h^2_\infty.
\label{ea.5}
\end{equation}

Following \cite{michel72}, we can combine \eqs{ea.1} and \eqref{ea.2} into the 
following differential equation
\begin{equation}
\begin{split}
 \bigg[1-\frac{a^2}{v^2}\bigg(1 & -\frac{2M}{r}  + v^2\bigg)\bigg] v\frac{\ud 
v}{\ud r} 
= \\ & -\frac{M}{r^2} + 2\,\frac{a^2}{r}\left(1-\frac{2M}{r} + v^2\right).
\label{ea.6}
\end{split}
\end{equation}
From \eq{ea.6} we obtain that the condition of having a critical point, i.e.~a 
radius $r=r_c$ at which both sides of this equation vanish simultaneously, 
translates into 
\begin{equation}
v_c^2=\frac{1}{2}\,\frac{M}{r_c},
\label{ea.7}
\end{equation}
and 
\begin{equation}
a_c^2=\frac{v_c^2}{1-3\,v_c^2}.
\label{ea.8}
\end{equation}

Substituting  \eqs{ea.7} and \eqref{ea.8} into \eq{ea.5} results in
\begin{equation}
n\,h_c^3 - h^2_\infty\left[(n+3)h_c-3\right]=0,
\label{ea.9}
\end{equation}
where $n=1/(\gamma-1)$. This polynomial has three real roots for $h_c$, but 
only one satisfies $h_c>1$ and thus has physical meaning. This root is given by
\begin{equation}
 h_c = 2\,h_\infty\,\sqrt{\frac{n+3}{3\,n}} 
\,\sin\left(\Psi+\frac{\pi}{6}\right), 
 \label{ea.10}
\end{equation}
where
\begin{equation}
 \cos(3\,\Psi) = \frac{3}{2\,n\,h_\infty}\left(\frac{n+3}{3\,n}\right)^{-3/2}.
\label{ea.11}
\end{equation}

We can now find an expression for the accretion rate in terms of $M$, the 
equation of state of the fluid, and its asymptotic conditions (expressed in 
terms of $\rho_\infty$ and $h_\infty$). Let us start  by substituting 
\eq{ea.7} into the continuity equation \eqref{ea.3}, which results in
\begin{equation}
\lambda = r_c^2\,\rho_c\,v_c = \frac{1}{4}\,M^2\rho_c\,v_c^{-3}.
\label{ea.12}
\end{equation}

Now, using \eqs{e3.1} and \eqref{e3.3}, we can rewrite the equation of state 
as 
\begin{equation}
\rho_c = \rho_\infty\left(\frac{h_c - 1}{h_\infty - 1}\right)^n,
\label{ea.13}
\end{equation}
on the other hand, by combining \eqs{e3.4}, \eqref{ea.8} and  \eqref{ea.10} 
we obtain
\begin{equation}
v_c^2 = \frac{h_c - 1}{(n+3)h_c - 3} = \frac{h^2_\infty(h_c - 1)}{n\,h^3_c} .
\label{ea.14}
\end{equation}
Then, substituting \eqs{ea.13} and \eqref{ea.14} into \eq{ea.12} we obtain
\begin{equation}
\dot{M}_\mathrm{M} = 4\pi\,\lambda = \pi\left[\frac{n^3h_c^9}{h^6_\infty}
\frac{(h_c - 1)^{2n-3}}{(h_\infty - 1)^{2n}}\right]^{1/2} M^2\rho_\infty.
\label{ea.15}
\end{equation}

Note that $\dot{M}_\mathrm{M}$ as expressed in \eq{ea.15} is given in terms of 
thermodynamical quantities measured asymptotically far away from the central 
object ($\rho_\infty$ and either $a_\infty$ or $ h _\infty$). In order to 
compare the results presented in Section~\ref{s2} with Michel's solution, one 
needs first to find the asymptotic values $\rho_\infty$ and  $a_\infty$ 
resulting in a solution such that $\rho|_\mathcal{R} = \rho_0$ and 
$a|_\mathcal{R} = a_0$.

\setcounter{equation}{0}
\setcounter{table}{0}
\renewcommand{\thefigure}{B.\arabic{figure}}
\renewcommand{\theequation}{B.\arabic{equation}}
\renewcommand{\thetable}{B.\arabic{table}}

\section{Validation of the numerical code}
\label{AB}

In this appendix, we present a series of standard tests in order to validate 
the results using {\it aztekas}. This section is complementary to the test 
already presented in Section~\ref{s3.1} as well as the appendix presented 
in~\citet{TA19}, where the code is tested using the relativistic spherical 
accretion problem~\citep{michel72}.

\subsection{One-dimensional shock tube tests}

\begin{figure*}
    \centering
\subfigure[Test 1]{
    \includegraphics[width=0.4\textwidth]{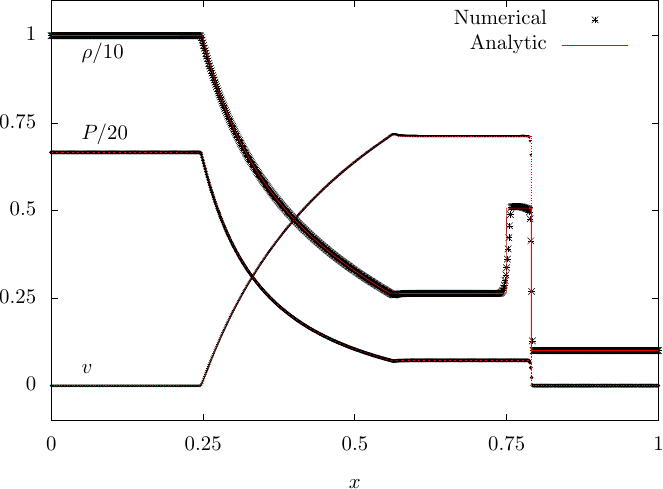}
}
\subfigure[Test 2]{
    \includegraphics[width=0.4\textwidth]{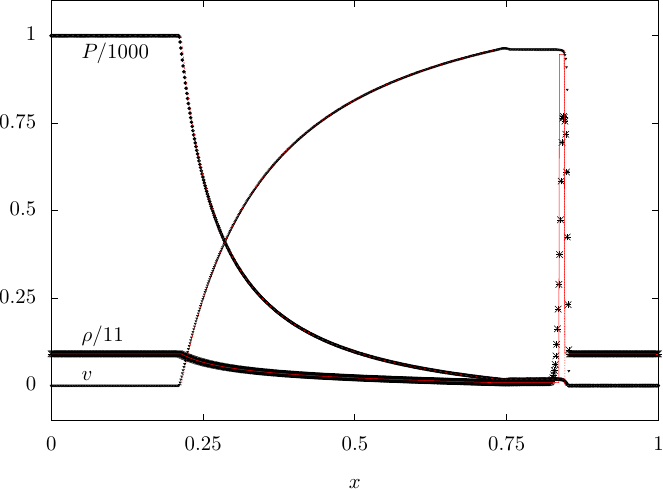}
}
\subfigure[Test 3]{
    \includegraphics[width=0.4\textwidth]{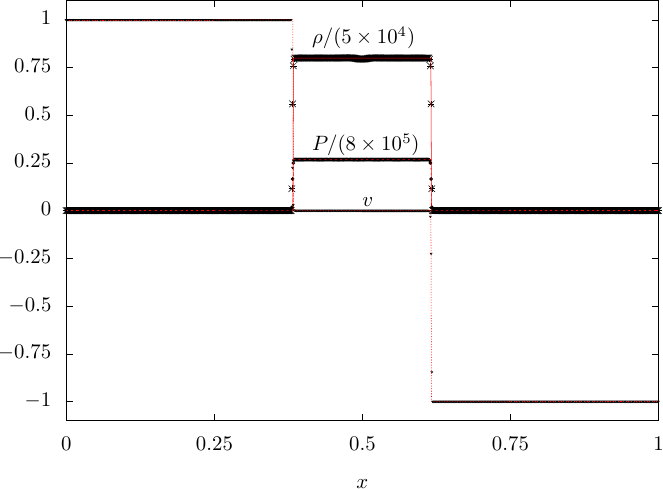}
}
\subfigure[Test 4]{
    \includegraphics[width=0.4\textwidth]{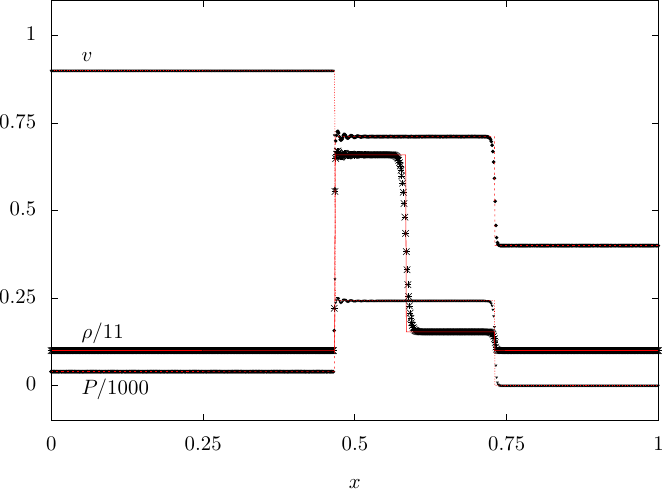}
}
\caption{Evolution at a time $t = 0.35$ of the density $\rho$, pressure $P$ and 
velocity $v$ for the four tests presented in Table~\ref{tab:tests}. All tests 
were performed using a resolution $N = 800$ with a Courant factor of 0.5.}
\label{fig:tests}
\end{figure*}

The one-dimensional shock tube test~\citep{sod1978} is a standard problem to 
solve for code validation, as it is easy to implement and the exact solution 
can be computed.  It consists of a perfect fluid at two different initial 
states with parameters $(\rho_L,P_L,v_L)$ and $(\rho_R,P_R,v_R)$ (where 
subscripts $L$ and $R$ refer to the left and right sides, respectively), 
separated by an interface at $x=x_0$. At $t=0$ the interface is removed and the 
two states are left to interact with each other. The evolution of this 
configuration depends only on the initial values and on the equation of state. 
Here we present a set of four one-dimensional shock tube tests, along with a 
two-dimensional version of the problem. For the former case, we compare the 
results with the analytic solution.

We reproduce four of the shock tube tests presented by~\citet{cafe}. The tests 
were performed in a Cartesian 1D domain $x \in [0,1]$ with a resolution $N = 
800$ and a Courant number of 0.5. The initial conditions of each test are 
presented in Table~\ref{tab:tests}. Test 1 and Test 2 correspond to a midly and 
strong relativistic blast wave explosion, respectively. The initial conditions 
for Test 3 produce a highly relativistic symmetric head-on stream collision, 
with a Lorentz factor $\Gamma = 1000$. Finally, Test 4 follows the evolution of 
a shock travelling at $v = 0.9$, as seen from the rest frame of the shock front.

\begin{table}[t]
    \centering
    \caption{Initial conditions for the left $(L)$ and right $(R)$ states of 
the set of one-dimensional shock tube tests.}
    \begin{tabular}{c|cccccc}
        \hline
                 & Test 1    & Test 2 & Test 3      & Test 4 \\
        \hline
        $\rho_L$ &  10       & 1      & 1           & 1      \\
        $P_L$    &  13.33    & 1000   & 0.001       & 1      \\
        $v_L$    &  0        & 0      &0.999999995  & 0.9    \\
        \hline 
        $\rho_R$ &  1        & 1      & 1           & 1      \\
        $P_R$    & $10^{-8}$ & 0.001  & 0.001       & 10     \\
        $v_R$    &  0        & 0      &-0.999999995 & 0      \\
        \hline
        $\gamma$ & 5/3       & 5/3    & 4/3         & 4/3    \\
        \hline
    \end{tabular}
    \label{tab:tests}
\end{table}

In Figure~\ref{fig:tests} we show the evolution of all four tests at $t=0.35$ 
and the comparison between the numerical simulations performed with 
\textit{aztekas} and the analytic solution. The latter was computed using a 
code written by~\citet{marti1999}. In the first two tests, a contact 
discontinuity and a rarefaction wave are formed. In Test 1, where the Lorentz 
factor is $\Gamma \approx 1$ the analytic solution is well resolved. In Test 2, 
where the Lorentz factor is $\Gamma \approx 6$, we obtain an overall 
satisfactory result although a higher resolution should lead to a better 
defined contact discontinuity. Test 3 shows the evolution of a strong head-on 
collision between two shock waves. A stationary high density, high pressure 
shell is formed and we see again a good match with the analytic solution. 
Finally, in Test 4, we can see the formation of a stationary contact 
discontinuity. In this case small oscillations are developed right behind the 
shock. This oscillations are damped with higher resolution. All the results 
presented here agree with previous works that use similar 
schemes~\citep[e.g.][]{lucas2004,decolle2012,cafe}.

\subsection{Two-dimensional Riemann problem}

\begin{figure}[t]
    \centering
    \includegraphics[width=0.45\textwidth]{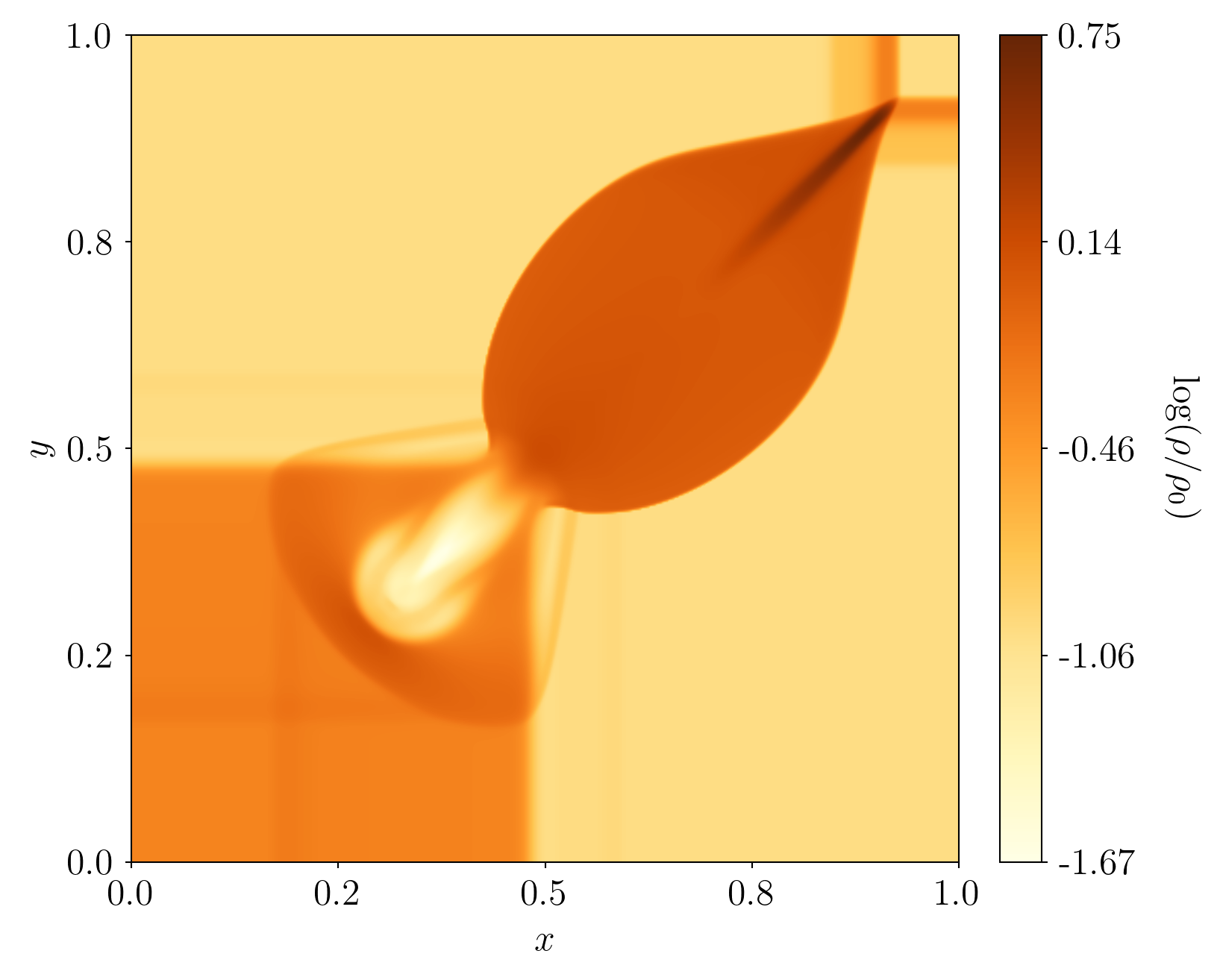}
    \caption{Logarithmic rest mass density colour map of the two-dimensional 
Riemann problem at a time $t=0.4$. The simulation was performed using a 
$400\times400$ resolution and a Courant factor of 0.25.}
    \label{fig:2dr}
\end{figure}

For this test, we follow closely the initial setup proposed 
by~\citet{delzanna2002} for the two-dimensional Riemann problem, which is the 
relativistic extension of the case presented by~\citep{lax1998}. The problem 
consists on a square domain subdivided into four regions with different initial 
states
\begin{equation}
    \footnotesize
    (\rho,P,v_x,v_y) = \left\lbrace
    \begin{split}
        (0.1,0.01,0,0) \quad &\mathrm{if} \quad x > 0.5 \text{,}\,\,\,y > 0.5, 
\\
        (0.1,1,0.99,0) \quad &\mathrm{if} \quad x \leq 0.5 \text{,}\,\,\, y > 
0.5, \\
        (0.5,1,0,0)\,\,\,\,\,\,\,\, \quad &\mathrm{if} \quad x > 0.5 
\text{,}\,\,\, y \leq 0.5, \\
        (0.1,1,0,0.99) \quad &\mathrm{if} \quad x \leq 0.5 \text{,}\,\,\, y 
\leq 0.5.
    \end{split}
    \right.
\end{equation}

The test was performed in a 2D Cartesian domain $(x,y) \in [0,1]\times[0,1]$ 
with a $400\times 400$ resolution and a Courant number of 0.25. In 
Figure~\ref{fig:2dr} we show the evolution of the rest mass density at a time 
$t=0.4$. The morphology of the solution shows a stationary high density contact 
discontinuity along the diagonal of the domain, and a jet-like structure 
propagating into the initially over-dense region. This results are 
qualitatively similar to the ones obtained 
by~\citet{delzanna2002,decolle2012,cafe}.

\bibliography{references}
\bibliographystyle{aasjournal}

\end{document}